\documentclass[twocolumn]{aastex61}

\graphicspath{{./}{figures/}}

\received{\today}
\revised{\today}
\accepted{\today}
\submitjournal{ApJ}

\shorttitle{[C{\sc ii}]-deficit in dwarf starbursts}
\shortauthors{Bisbas et al.}

\defcitealias{Lahe20}{L20}

\begin{document}

\title{The origin of the [CII]-deficit in a simulated dwarf galaxies starburst}

\correspondingauthor{Thomas G. Bisbas}
          
          \email{bisbas@ph1.uni-koeln.de}

\author[0000-0003-2733-4580]{Thomas G. Bisbas}
\affil{I. Physikalisches Institut, Universit\"at zu K\"oln, Z\"ulpicher Stra\ss e 77, K\"oln, Germany}
\affil{Department of Physics, Aristotle University of Thessaloniki, GR-54124 Thessaloniki, Greece}

\author{Stefanie Walch}
\affil{I. Physikalisches Institut, Universit\"at zu K\"oln, Z\"ulpicher Stra\ss e 77, K\"oln, Germany}

\author{Thorsten Naab}
\affil{Max Planck Institut f\"ur Astrophysik, Karl-Schwarzschild-Str. 1, D-85741 Garching, Germany}

\author{Natalia Lah\'en}
\affil{Max Planck Institut f\"ur Astrophysik, Karl-Schwarzschild-Str. 1, D-85741 Garching, Germany}

\author{Rodrigo Herrera-Camus}
\affil{Departamento de Astronomía, Universidad de Concepción, Barrio Universitario, Concepción, Chile}

\author{Ulrich P. Steinwandel}
\affil{Center for Computational Astrophysics, Flatiron Institute, 162 5th Avenue, New York, NY 10010}

\author{Constantina M. Fotopoulou}
\affil{Max Planck Institut f\"ur Astrophysik, Karl-Schwarzschild-Str. 1, D-85741 Garching, Germany}

\author{Chia-Yu Hu}
\affil{Max-Planck-Institut f\"ur Extraterrestrische Physik, Giessenbachstrasse 1, D-85748 Garching, Germany}

\author{Peter H. Johansson}
\affil{Department of Physics, University of Helsinki, Gustaf H\"allstr\"omin katu 2, FI-00014 Helsinki, Finland}

\begin{abstract}
We present [C{\sc ii}] synthetic observations of smoothed particle hydrodynamics (SPH) simulations of a dwarf galaxy merger. The merging process varies the star-formation rate by more than three orders of magnitude. Several star clusters are formed, the feedback of which disperses and unbinds the dense gas through expanding H{\sc ii} regions and supernova (SN) explosions. For galaxies with properties similar to the modelled ones, we find that the [C{\sc ii}] emission remains optically thin throughout the merging process. We identify the Warm Neutral Medium ($3<\log T_{\rm gas}<4$ with $\chi_{\rm HI}>2\chi_{\rm H2}$) to be the primary source of [C{\sc ii}] emission ($\sim58\%$ contribution), although at stages when the H{\sc ii} regions are young and dense (during star cluster formation or SNe in the form of ionized bubbles) they can contribute $\gtrsim50\%$ to the total [C{\sc ii}] emission. We find that the [C{\sc ii}]/FIR ratio decreases due to thermal saturation of the [C{\sc ii}] emission caused by strong FUV radiation fields emitted by the massive star clusters, leading to a [C{\sc ii}]-deficit medium. We investigate the [C{\sc ii}]-SFR relation and find an approximately linear correlation which agrees well with observations, particularly those from the Dwarf Galaxy Survey. Our simulation reproduces the observed trends of [C{\sc ii}]/FIR versus $\Sigma_{\rm SFR}$ and $\Sigma_{\rm FIR}$, and it agrees well with the Kennicutt relation of SFR-FIR luminosity. We propose that local peaks of [C{\sc ii}] in resolved observations may provide evidence for ongoing massive cluster formation.
\end{abstract}


\section{Introduction} 
\label{sec:intro}

The star-formation rate (SFR) is of fundamental importance for understanding the cyclic process of global star formation in galaxies across the epochs \citep[see][for a review]{Mada14}. Measuring it can reveal the properties and the evolutionary stages of the observed interstellar medium (ISM). \citet{Schm59} and \citet{Kenn98} were the first to find a strong correlation between the SFR per unit area and the gas surface density, a relation frequently referred to as ``Schmidt-Kennicutt relation". Since then, various methods based on continuum bands and optical/near-infrared (IR) emission lines have been used to measure SFR in different systems \citep[see][for a review]{Kenn12}. Recent attempts using fine-structure lines such as [O{\sc i}] at $63\,\mu$m and [O{\sc iii}] at $88\,\mu$m \citep[e.g.][]{Hunt01,Brau08,DeLo14,Olse17}, as well as H$\alpha$, UV and IR \citep[e.g.][]{Shiv15} have shown good correlations with the SFR. The far-IR fine-structure transition of [C{\sc ii}] $^2P_{3/2}-{^2P}_{1/2}$ at a rest frame wavelength of $157.7\,\mu$m (hereafter referred to simply as [C{\sc ii}]) is also widely used as a promising diagnostic of the SFR \citep[see][for early attempts]{Stac91,Bose02}. The Atacama Millimeter Array (ALMA) is able to provide unprecedented resolution of high-redshift ($z>1$) observations in this line (using Bands $5-10$, depending on the redshift), opening an entirely new window in the study of the Early Universe ISM.

[C{\sc ii}] is one of the brightest lines originating from star-forming galaxies \citep[e.g.][]{Stac91,Brau08}. The ionization potential of atomic carbon is $11.3\,{\rm eV}$, slightly lower than the ionization potential of hydrogen ($13.6\,{\rm eV}$). Under typical ISM environmental conditions, the [C{\sc ii}] emission line is a result of the interaction between the ISM gas and FUV photons. In general, the emission of [C{\sc ii}] represents ${\sim}\,10^{-3}$ of the total far-infrared (FIR) continuum emission. Furthermore, it is also found to be associated with the outer shells of H$_2$-rich clouds, where star-formation takes place. Thus, [C{\sc ii}] plays a very important role in photodissociation regions (PDRs) as a coolant, particularly at low visual extinctions \citep{Holl99,Wolf03}. It has an upper-state energy of $h\nu/k_{\rm B}\sim91\,{\rm K}$ and its critical density spans approximately three orders of magnitude depending on the temperature \citep{Gold12}. [C{\sc ii}] may, therefore, arise from different phases of the ISM, such as H{\sc ii}-regions, PDRs and cold molecular gas \citep{Velu14,Abdu17,Crox17,Accu17,Laga18,Ferr19,Corm19} depending on the environmental parameters, such as the intensity of the FUV radiation, metallicity, the cosmic-ray ionization rate \citep{Bisb15,Bisb17,Bisb19,Bisb21} and the intensity of X-rays \citep{Mack19}. Interestingly, the studies of \citet{Velu14} and \citet{Accu17} found that $\sim60-85\%$ of the total [C{\sc ii}] emission arises from the molecular gas phase, thus naturally explaining its correlation with SFR \citep[see also][]{Madd20}. However, numerical simulations of \citet{Fran18} show that if the cloud is young enough, its emission in [C{\sc ii}] arising from the molecular phase may be smaller than 20 per cent. In this regard, this fine-structure line may not be a good tracer for the CO-dark\footnote{The term `CO-dark' gas has been introduced by \citet{vDis92}} H$_2$-gas.

The [C{\sc ii}]/FIR luminosity ratio is observed to decrease with increasing infrared luminosity \citep{Malh97,Malh01,Luhm98,Luhm03,Case04}. The origin of the so called `[C{\sc ii}]-deficit' is still being investigated despite numerous efforts proposing a variety of mechanisms behind it \citep[e.g.][]{Malh01,Luhm03,Stac10,Grac11,Sarg12}. 
Suggestions include optically thick [C{\sc ii}] emission in large columns of dust, conversion of singly (C$^+$)\footnote{The [C{\sc ii}] notation refers to the emission of the line whereas the C$^+$ notation to the actual species and/or its abundance.} to doubly (C$^{2+}$) ionized carbon, and fine-structure lines (e.g. [O{\sc i}]) overcoming [C{\sc ii}] as coolants. \citet{Muno16} studied how strong FUV radiation fields can drive a low ratio of [C{\sc ii}]/FIR due to thermal saturation of the [C{\sc ii}] emission \citep[see also][]{Kauf99}. This mechanism has been recently confirmed observationally by \citet{Ryba19} and, as we will see later, it is also in accordance with our simulations for which we find a decreasing [C{\sc ii}]/FIR as the surface densities of FIR and SFR increase.

\citet{Nara17} provided a theoretical model suggesting that the cloud structure in galaxies with increasing SFRs, and hence increasing gas surface densities, is responsible for the [C{\sc ii}]/FIR ratio decrease. Using a large sample of ${\sim}\,15,000$ resolved regions, \citet{Smit17} was able to show that even extragalactic regions of a few hundred of parsecs in size appear to be [C{\sc ii}]-deficient. However, the effect is more prominent in the high-redshift Universe where distant and, thus, [C{\sc ii}]-faint sources may emit only $\sim10\%$ of the expected [C{\sc ii}] based on their observed FIR luminosity.

\begin{table*}[]
    \centering
    \begin{tabular}{c|c|c|c|l|l}
    $m$  & $b$ & $m_{\Sigma}$ & $b_{\Sigma}$ & Type of objects & Reference \\ \hline\hline 

    0.80 & $-5.73$ & 0.93 & $-6.99$ & Dwarf galaxies (DGS)& \citet{DeLo14}  \\ 
    1.01 & $-6.99$ & -- & -- & Various types of galaxies &\citet{DeLo14} \\ 
    0.98 & $-7.67$ & 1.13 & $-8.47$ & Normal star-forming galaxies & \citet{Herr15} \\
    0.98 & $-6.89$ & 0.99 & $-7.19$  & Milky Way clouds & \citet{Pine14}\\
    0.96 & $-7.22$ & 1.04 & $-7.81$ & Normal local galaxies & \citet{Sutt19} \\

    \hline
    \end{tabular}
    \caption{Summary of constants $m$, $b$, $m_{\Sigma}$ and $b_{\Sigma}$ considered here characterizing the best-fitting relation given by Eqns.~(\ref{eqn:ciisfr}) and (\ref{eqn:Sciisfr}) for different types of objects.}
    \label{tab:obs}
\end{table*}

In a series of hydrodynamical simulations with a resolution of $4\,{\rm M}_{\odot}$ per gas particle, \citet{Hu16,Hu17,Hu19} examined the global star formation process and how supernovae (SNe) affect the SFR in dwarf galaxies, as well as the underlying ISM microphysics including heating/cooling mechanisms and dust sputtering. Follow-up work by the {\sc griffin}\footnote{Galaxy Realizations Including Feedback From INdividual massive stars; https://wwwmpa.mpa-garching.mpg.de/$\sim$naab/griffin-project/index.html} collaboration (\citealt{Lahe20}; hereafer `\citetalias{Lahe20}') proposed that dwarf galaxy mergers may result in a significant population of star clusters. In particular, during the merging process the SFR may increase up to three orders of magnitude and can form clusters in the range of $10^2-10^{6}\,{\rm M}_{\odot}$. Such a large variation of SFRs in the dynamical evolution provides an interesting set of three-dimensional morphological ISM distributions, including feedback, which can in turn reveal insights on the origin of the [C{\sc ii}]-SFR correlation and the [C{\sc ii}]-deficit.

The focus of this work is to perform [C{\sc ii}] synthetic observations \citep[see][for a review]{Hawo18} of the {\sc griffin} dwarf galaxy merger simulations presented in \citetalias{Lahe20} and compare the results against existing observations. Apart from the [C{\sc ii}]/FIR ratio, another key study of this project is the [C{\sc ii}]-SFR relationship. In general, observational \citep{DeLo11,DeLo14,Pine14,Herr15,Herr18,Zane18,Sutt19} and numerical \citep{Olse15,Olse17,Vall15,Lupi20} studies, find a [C{\sc ii}]-SFR relation of the form:
\begin{eqnarray}
\label{eqn:ciisfr}
\log_{10}\frac{\rm SFR}{[\rm M_{\odot} \,yr^{-1}]}=m\log_{10}\frac{L_{\rm CII}}{[\rm L_{\odot}]}+b
\end{eqnarray}
where $0.8\lesssim m \lesssim1.2$ and $-8\lesssim b \lesssim-5$ depending on the type and redshift of the galaxy. When correlating the star-formation rate surface density ($\Sigma_{\rm SFR}$) with the surface [C{\sc ii}] luminosity ($\Sigma_{\rm CII}$), the above relation takes the form:
\begin{eqnarray}
\label{eqn:Sciisfr}
\log_{10}\frac{\Sigma_{\rm SFR}}{[\rm M_{\odot} \,yr^{-1}\,kpc^{-2}]}=&&m_{\Sigma}\log_{10}\frac{\Sigma_{\rm CII}}{[\rm L_{\odot}\,kpc^{-2}]}\nonumber\\ &+&b_{\Sigma}.
\end{eqnarray}
Table~\ref{tab:obs} provides a summary of the $m$, $b$, $m_{\Sigma}$ and $b_{\Sigma}$ values used in this work. 

This paper is organized as follows. 
Section~\ref{sec:sims} gives a description of the selected snapshots from the {\sc griffin} SPH simulations and the post-processing strategy.
Section~\ref{sec:time} discusses how the C{\sc ii} luminosity and the SFR vary in time throughout the merging process. Section~\ref{sec:origin} studies the origin of the [C{\sc ii}] emission and 
Section~\ref{sec:ciidef} its relation with the FIR emission and how the feedback from massive clusters leads to a [C{\sc ii}]-deficient medium. Finally, in
Section~\ref{sec:discussion} we compare our results with observations and examine the relation between SFR and FIR as well as the [C{\sc ii}] and SFR. 
We conclude in Section~\ref{sec:conclusions}.

\section{Description of simulations and the post-processing technique}
\label{sec:sims}
 
The hydrodynamical simulations of the dwarf galaxy merger are fully described in \citetalias{Lahe20}. These are Smoothed Particle Hydrodynamics simulations using the {\sc SPHGal} code presented in \citet{Hu14,Hu16,Hu17}, which is a modified version of {\sc gadget-3} \citep{Spri05} adopting a modern formulation of SPH that includes time-dependent artificial diffusion (viscosity and conduction) to improve SPH fluid-mixing behaviour. 
The calculations include a non-equilibrium model for cooling and chemistry that directly integrates the rate equations of H$_2$, H$^+$ and CO, while obtaining non-equilibrium abundances for H, C$^+$, O, and free electrons from residual conservation laws \citep{Nels97,Glov07}. They also take into account metal line cooling from 12 different metal species (H, He, N, C, O, Si, Mg, Fe, S, Ca, Ne and Zn) based on the cooling tables of \citet{Wier09} as implemented in \citet{Aume13}. For reference and for the $Z=0.1\,{\rm Z}_{\odot}$ environmental condition, the initial abundances of C$^+$ and O relative to hydrogen were $2.46\times10^{-5}$ and $4.9\times10^{-5}$, respectively (see \citetalias{Lahe20} for details). 

In addition, routines treating the interstellar radiation field, as well as stellar feedback in terms of photoionization, photoelectric heating and SNe are included. Ionization due to collisions is also treated as described in \citet{Hu17}. The treatment of H{\sc ii} regions has been described in \citet{Hu17} and it is a Str{\"o}mgren-type approximation for photoionization as discussed in \citet{Hopk12}. This approach was chosen so as to reduce the computational cost as opposed to a more detailed radiative transfer approach. We note that the adopted chemistry model does not account for the conversion of C{\sc ii} to C{\sc iii} which may somewhat overestimate the calculated C{\sc ii} luminosity. However, as we will see later, the contribution of H{\sc ii} regions to the total C{\sc ii} luminosity is small and thus a more detailed approach shall not alter the results presented in this work. The dynamical expansion of H{\sc ii} regions has been also benchmarked against the results of the STARBENCH workshop \citep{Bisb15a} with reasonable agreement.

With $4\,{\rm M}_{\odot}$ per SPH particle, the simulation resolves the Sedov-Taylor stage of individual SN-remnants for $>90\%$ of the ambient SN-densities \citep{Hu17,Hu19,Stei20}. Prior to the collision, the two dwarf galaxies are identical with virial masses of $M_{\rm vir}=2\times10^{10}\,{\rm M}_{\odot}$, virial radii of $r_{\rm vir}=44\,{\rm kpc}$ and are composed of a dark matter halo and a gas-rich disk with a rotationally supported (old) stellar population. The two galaxies are set on parabolic orbits with a pericentric distance of $1.46\,{\rm kpc}$ and an initial separation of $5\,{\rm kpc}$. The collision is not edge-on but includes an inclination similar to the Antennae galaxies merger, as described in \citet{Lahe19}.

Under the optically thin assumption (see \S\ref{sec:time}), the total C{\sc ii} luminosity is given by the expression
\begin{eqnarray}
\label{eqn:Lcii_sum}
L_{\rm CII} = 2.61\times10^{-34}\sum_{i=1}^{N_{\rm SPH}}\frac{\Lambda_{\rm CII,i}}{n_{{\rm H},i}}\frac{m_{{\rm SPH},i}}{m_{\rm p}}\,[{\rm L}_{\odot}],
\end{eqnarray}
where $\Lambda_{\rm CII}$ is the C{\sc ii} cooling function of the $i-$th SPH particle\footnote{Each `SPH particle' covers a spherical volume defined by the number of $N_{\rm neighb}$ neighbouring particles.} (in units of $\rm erg\,s^{-1}\,cm^{-3}$), $m_{\rm SPH}$ is its corresponding mass, $m_{\rm p}$ is the proton mass and the summation is over all particles within the volume of interest. We construct luminosity maps at a resolution of $1024^2$ uniform pixels, where we project the SPH particles. The luminosity of each pixel is then a direct summation of the SPH particles along the line-of-sight using the above equation.

\begin{figure*}
    \centering
    \includegraphics[width=0.99\textwidth]{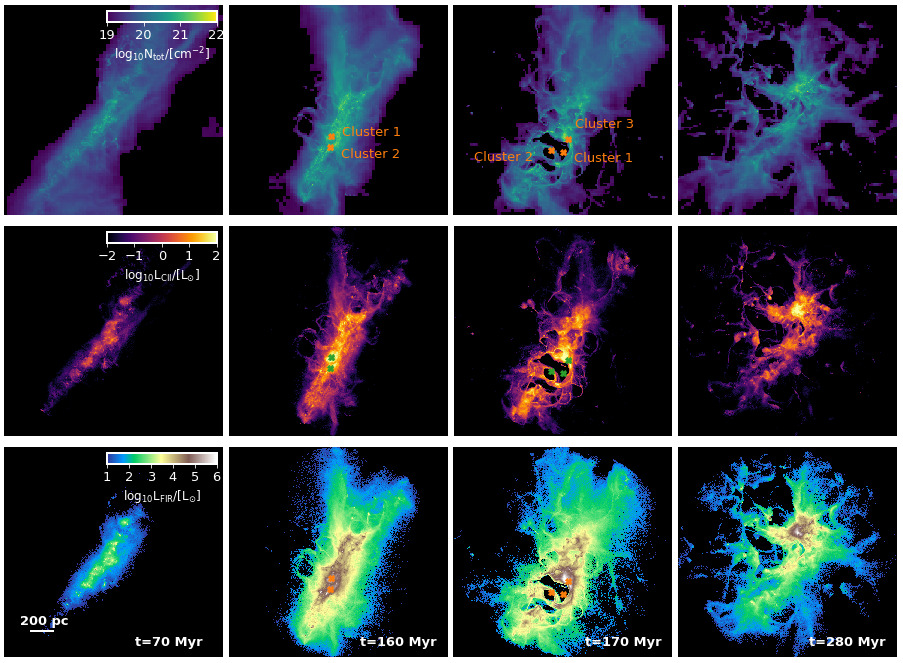}
    \caption{Snapshots showing the total gas column density (top row), the C{\sc ii} luminosity (middle row), and the FIR luminosity (bottom row) for four different snapshots taken at 70, 160, 170 and 280~Myr. Each side has a size of 2~kpc. The three most massive clusters are formed during the merging process, indicated in the $t=160$ and $170\,{\rm Myr}$ panels, and create expanding H{\sc ii} regions. The merger becomes bright in [C{\sc ii}] emission for $t\gtrsim150\,{\rm Myr}$ when SFR~$\gtrsim10^{-3}\,{\rm M}_{\odot}\,{\rm yr}^{-1}$ (see also Fig.~\ref{fig:time}). Similarly, the FIR luminosity increases during the second merger and remains high thereafter.}
    \label{fig:snaps}
\end{figure*}

We calculate the total far-infrared (FIR) emissivity by adopting the dust cooling rate which is given by the expression:
\begin{eqnarray}
\Lambda_{\rm dust}=4\pi \rho\int_0^{\infty}B_{\nu}(T_{\rm d})\kappa_{\nu}d\nu,
\end{eqnarray}
where $\rho$ is the gas density, $B_{\nu}(T_{\rm d})$ is the Planck function at frequency $\nu$, $T_{\rm d}$ is the dust temperature and $\kappa_{\nu}$ is the dust opacity. \citet{Glov12} fit this relation using the expression \citep[see also][]{Hu17}:
\begin{eqnarray}
\label{eqn:ld}
\Lambda_{\rm dust}=4.68\times10^{-31}D'T_{\rm d}^6n_{\rm H}\,[{\rm erg}\,{\rm s}^{-1}\,{\rm cm^{-3}}],
\end{eqnarray}
where $D'$ is the dust-to-gas mass ratio relative to the solar value\footnote{For solar metallicity, $D'=1.0$}. Here, we set $D'=0.1$ since the modelled dwarf galaxies have a metallicity of $Z=0.1\,{\rm Z}_{\odot}$. For the purposes of this work this linear relation between dust-to-gas ratio and metallicity is generally a good assumption, although metal-poor systems with metallicities lower than the one examined here may not follow such a relation \citep{Herr12,Remy14}. Equation~\ref{eqn:ld} is valid for $5{<}\,T_{\rm d}{<}\,100\,{\rm K}$. The ${\propto}\,T_{\rm d}^6$ dependency arises from the Stefan-Boltzmann law and the opacity term. The total FIR luminosity as well as the FIR luminosity maps, are then constructed as described above for the C{\sc ii} luminosity.

\begin{figure*}
    \centering
    \includegraphics[width=0.92\textwidth]{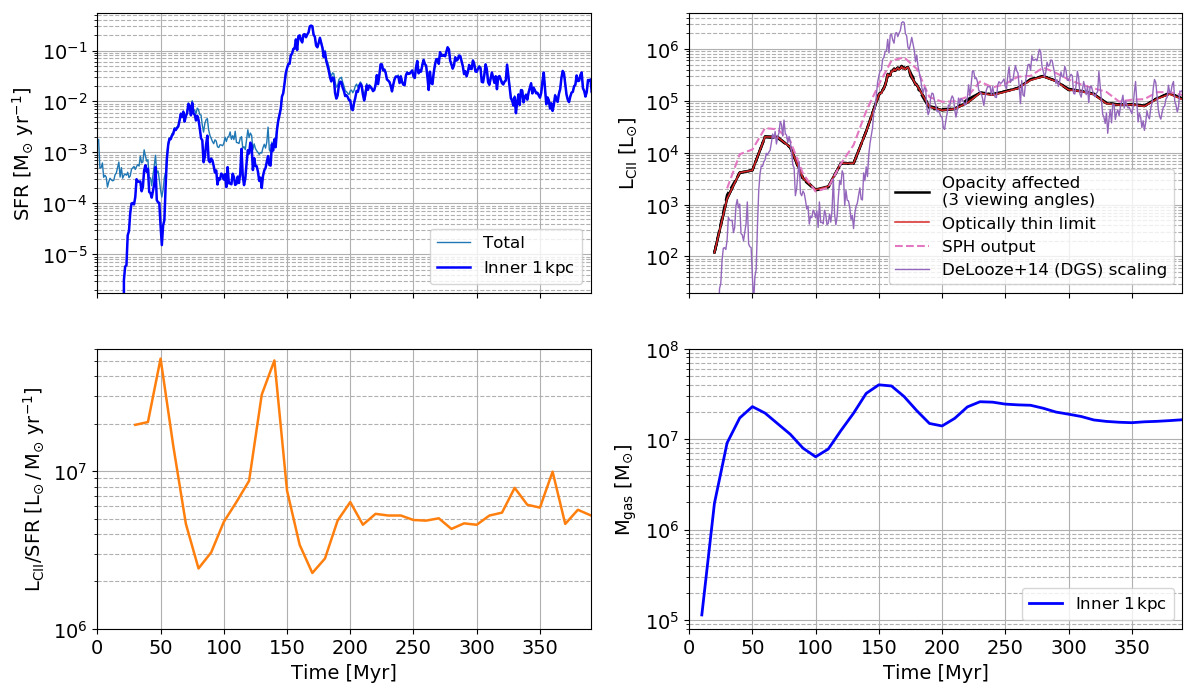}
    \caption{{\it Top left panel:} Star-formation rate versus time (at a resolution of $\Delta t=1$~Myr) calculated for the entire computational domain (`Total'; thin light blue line) and for the inner $1\,{\rm kpc}$ radius (`Inner'; thick dark blue line). The peaks at ${\sim}80$ and ${\sim}160\,{\rm Myr}$ correspond to the first and second encounter of the merger, respectively. 
    {\it Top right panel:} [C{\sc ii}] luminosity, $L_{\rm CII}$, versus time (at a resolution of $\Delta t=10$~Myr). The {\sc radmc-3d} opacity-affected calculations (black lines) for three different viewing angles and the corresponding one for  the optically-thin calculation (red line) are overplotted. The lines are indistinguishable implying that the optically thin emission is an excellent approximation for $L_{\rm CII}$. The dashed magenta line is the $L_{\rm CII}$ derived directly from the SPH particles using Eqn.~\ref{eqn:Lcii_sum}. The thin solid purple line shows $L_{\rm CII}$ scaled using Eqn.~\ref{eqn:ciisfr} for the \citet{DeLo14} DGS survey. 
    {\it Bottom left panel:} The $L_{\rm CII}$/SFR ratio versus time ($L_{\rm CII}$ derived directly from the SPH particles and SFR calculated for the inner 1~kpc and averaged over the preceding $5\,{\rm Myr}$). The ratio decreases during both merging processes and fluctuates within one order of magnitude, contrary to the three orders of magnitude span of both $L_{\rm CII}$ and SFR. 
    {\it Bottom right panel:} The gas mass versus time (at a resolution of $\Delta t=10$~Myr) for the inner 1~kpc. }
    \label{fig:time}
\end{figure*}

In case we consider velocity-resolved emission properties and unless stated otherwise, we impose a lower observational limit for the [C{\sc ii}] 
luminosity of $\rm L_{\rm CII}=0.5\,{\rm L}_{\odot}$, corresponding to a velocity integrated emission of $\rm W(CII)\simeq0.6\,K\,km/s$ (see Eqn.~\ref{eqn:Lcii} for a resolution of $1024^2$ and Appendix~\ref{app:surface} for the effect of using a different lower limit). This is a reasonable assumption considering the sensitivity of instruments \citep[see also][]{Fran18}. The FIR luminosity considered in this analysis, corresponds to the pixels that satisfy the aforementioned C{\sc ii} observational criterion. In a similar way, the surface, $\Sigma$, is estimated by the area covered from the above number of pixels.

There are two merging events; a first passage encounter occurring at $t\sim80\,{\rm Myr}$ and a second encounter leading to a final coalescence occurring at $t\sim170\,{\rm Myr}$. We post-process snapshots from $t=10\,{\rm Myr}$ to $t=390\,{\rm Myr}$ with a $10\,{\rm Myr}$ step. We therefore post-process a total of 39 snapshots. The SFR is calculated from the simulation snapshots following the methodology of \citet{Hu16}. This methodology is a stochastic star formation approach where the local SFR is $\epsilon_{\rm sf}\rho / t_{\rm ff}$, where $\rho$ is the gas density, $t_{\rm ff}$ is the free-fall time, and $\epsilon_{\rm sf} = 0.02$ is the star formation efficiency. The SFR quantity used in the present work is an average over the past 1~Myr.

The top row of Fig.~\ref{fig:snaps} shows the total gas column density, $N_{\rm tot}$, the middle row the corresponding [C{\sc ii}] luminosity and the bottom row the FIR luminosity in four different snapshots. These are (from left to right) during the first encounter at $t=70\,{\rm Myr}$, during the second encounter at $t=160$ and $t=170\,{\rm Myr}$, and after the gas settling in the central part at $t=280\,{\rm Myr}$. During the first encounter, the bar-like structure becomes bright in [C{\sc ii}] only in its central part where the gas surface density becomes high enough $N_{\rm tot}{\simeq}\,10^{21}\,{\rm cm}^{-2}$. During the second encounter, clusters are formed in the dense parts. They produce ionizing radiation creating H{\sc ii}-regions. These, in turn, are responsible for the bubble-like features seen at $t=170\,{\rm Myr}$. The positions of the three most massive clusters formed are indicated in these snapshots. The masses of the clusters are $1.6$, $1.2$, and $7.9\,{\times}\,10^5\,{\rm M}_{\odot}$ for the first, second and third most massive cluster, respectively (\citetalias{Lahe20}). In the final phase shown in Fig.~\ref{fig:snaps}, the two disks merge and enhance feedback from H{\sc ii}-regions. Subsequent SNe disperse and unbind the gas, creating the irregular shape seen at $t=280\,{\rm Myr}$.

\section{Time evolution of ionized carbon luminosity and Star Formation Rate}
\label{sec:time}

\begin{figure*}
    \centering
    \includegraphics[width=0.98\textwidth]{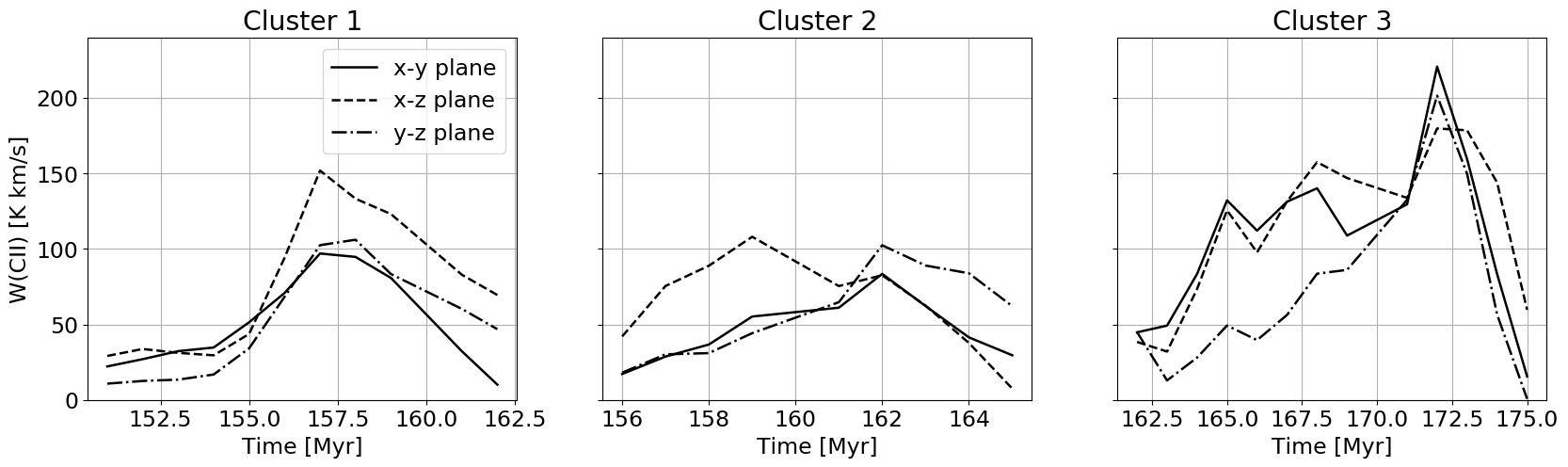}
    \caption{Time-evolution of the [C{\sc ii}] velocity integrated emission for the three main clusters (left to right). The velocity integrated emission (produced here with {\sc radmc-3d} see Appendix~\ref{app:radmc3d}) is an average over 25 pixels ($1.5\times10^3\,{\rm pc}^2$) centered on the position of each cluster. Each different line-type corresponds to a different viewing angle. The solid line is for the viewing angle of Fig.~\ref{fig:snaps}. As can be seen, there is good agreement of the [C{\sc ii}] trends as a function of time regardless of the viewing angle. The differences observed in each viewing angle are not arising from optical depth effects but rather due to the different projections of mass covered by the above area.}
    \label{fig:clusters}
\end{figure*}

The time evolution of the SFR, $L_{\rm CII}$ and $L_{\rm CII}$/SFR for all gas mass is shown in Fig.~\ref{fig:time}. The top left panel shows the SFR calculated for the entire computational domain (thin light blue line) as well as for the inner $1\,{\rm kpc}$ (thick dark blue line). The latter SFR is the one we will use throughout this work. The two SFRs are in excellent agreement when the dwarf galaxies experience an encounter. We note that for the calculation of SFR in this panel, we use the outputs of the simulation in time intervals of $\Delta t=1\,{\rm Myr}$.

As described in \citetalias{Lahe20} (see their Fig.~2), the first pericentric passage occurs at ${\sim}\,50\,{\rm Myr}$ and the first apocenter at ${\sim}\,80\,{\rm Myr}$. During that period of time, a tidal bridge forms connecting the two galaxies which results in an approximately two orders of magnitude increase of the SFR, from ${\sim}\,10^{-4}$ to ${\sim}\,10^{-2}\,{\rm M}_{\odot}\,{\rm yr}^{-1}$. The second and much stronger encounter occurs between ${\sim}\,150$ to ${\sim}\,180\,{\rm Myr}$ with the SFR peaking at ${\sim}\,160\,{\rm Myr}$. This is the starburst phase where multiple clumpy star formation regions exist. During this second period, the SFR reaches values as high as $0.2-0.3\,{\rm M}_{\odot}{\rm yr}^{-1}$, corresponding to a mini-starburst. 
Earlier works by \citet{Hu16,Hu17} showed that in an isolated dwarf galaxy with properties similar to those modelled here, the SFR is approximately $10^{-4}-10^{-3}\,{\rm M}_{\odot}{\rm yr}^{-1}$ and relatively constant. This in turn means that throughout the merging process of such dwarf galaxies, the SFR can vary between two and three orders of magnitude depending on the evolutionary stage.

It is interesting to explore whether or not a potential assumption of optically thin [C{\sc ii}] emission is valid. We do this since [$^{13}$C{\sc ii}] emission of the Large Magellanic Cloud studied by \citet{Okad19} shows that [C{\sc ii}] may become optically thick, having implications for the extragalactic observations of this line and, thus, the obtained [C{\sc ii}]-SFR relation. For this investigation, we perform additional calculations with the radiative transfer code {\sc radmc-3d} (see Appendix~\ref{app:radmc3d}). In the top right panel of Fig.~\ref{fig:time}, we plot with black lines the $L_{\rm CII}$  calculated with {\sc radmc-3d} versus time for three different viewing angles ($x-y$, $x-z$, $y-z$ planes). On top of these three lines, we plot with red solid line the corresponding {\sc radmc-3d} calculations in the optically thin limit. As can be seen, all aforementioned lines are indistinguishable showing that in these simulations [C{\sc ii}] can be very well approximated as optically thin. Furthermore, we plot with dashed magenta line the $L_{\rm CII}$ calculated using Eqn.~\ref{eqn:Lcii_sum} which is in excellent agreement with the {\sc radmc-3d} results. Finally, in this panel we also plot with a thin solid purple line the $L_{\rm CII}$ value given from Eqn.~\ref{eqn:ciisfr} for $m\,{=}\,0.80$ and $b\,{=}\,-5.73$ corresponding to the DGS of \citet{DeLo14}. As we describe in \S\ref{ssec:sfirssfr}, our simulations show an excellent agreement with these observations which best represent our modelled galaxies.

The modelled galaxies are low-mass with low metallicity so any opacity effects are negligible. We actually verified this by performing {\sc radmc-3d} calculations. However, \citet{Fran18} showed that during molecular cloud formation at solar metallicity, [C{\sc ii}] can become quickly optically thick. On the other hand, \citet{Bisb21} showed that metal-poor clouds may remain optically thin for H$_2$ column densities up to ${\sim}10^{23}\,{\rm cm}^{-2}$ while the presence of strong FUV intensities can positively contribute to the increment of the [C{\sc ii}] optical depth. Such conditions are exceptional for the modelled dwarf galaxies, thus [C{\sc ii}] may remain always optically thin. We argue that the [C{\sc ii}] emission of systems with similar properties to the simulated galaxies is in general optically thin. However, we cannot unambiguously demonstrate that larger systems and especially galaxies with metallicities close to solar will remain [C{\sc ii}] optically thin.

The first encounter results in an increase of $L_{\rm CII}$ spanning approximately two orders of magnitude, reaching ${\sim}\,2\times10^4\,{\rm L}_{\odot}$ at ${\sim}\,70\,{\rm Myr}$. When the dwarf galaxies reach the apocenter at ${\sim}\,80\,{\rm Myr}$, the column density decreases resulting in a decrease in SFR and $L_{\rm CII}$. The second, stronger, encounter results in a much more prominent increase in $L_{\rm CII}$, reaching values ${\sim}\,5\times10^5\,{\rm L}_{\odot}$. As described in \citetalias{Lahe20}, SN-feedback from the clusters disperses the dense distribution of gas. This leads to a decrease of $L_{\rm CII}$ at ${\sim}\,180\,{\rm Myr}$. However, for times ${>}\,200\,{\rm Myr}$, $L_{\rm CII}$ fluctuates following the trend of SFR due to the settling of the gas in the central region and the associated feedback.

The bottom left panel of Fig.~\ref{fig:time} shows how the $L_{\rm CII}$/SFR ratio evolves in time. In this panel, we have considered an average SFR value over the preceding $5\,{\rm Myr}$, every $10\,{\rm Myr}$. As can be seen, this ratio remains approximately constant at a value of ${\sim}\,3-6\times10^6\,{\rm L_{\odot}/M_{\odot}\,yr^{-1}}$ for $t\,{>}\,200\,\rm Myr$. Overall, the ratio does not strongly vary when compared to the fluctuations of both $L_{\rm CII}$ and SFR that span more than three orders of magnitude throughout the evolution. The $L_{\rm CII}$/SFR ratio decreases during the first and especially during the second encounter. This relatively small fluctuation of this ratio compared to the corresponding one observed in both SFR and $L_{\rm CII}$ individually, indicates that $L_{\rm CII}$ is a good tracer for estimating the SFR.

The bottom right panel of Fig.~\ref{fig:time} shows the total gas mass versus time at an interval of $\Delta t=10\,{\rm Myr}$. After $t>50\,{\rm Myr}$, the average gas mass is $\sim2\times10^7\,{\rm M}_{\odot}$, which is $\sim1/4$ of the total gas mass of the simulation setup described in \citetalias{Lahe20}.

Figure~\ref{fig:clusters} shows the time evolution of the [C{\sc ii}] velocity integrated emission for the regions within which the three most massive clusters form. The plotted velocity integrated emission (here produced with {\sc radmc-3d}; see Appendix~\ref{app:radmc3d}) is an average over 25 pixels which are centered around each cluster position shown in Fig.~\ref{fig:snaps}. The linear size of each of these pixels is approximately $7.8\,{\rm pc}$, thus covering an area of $\sim61\,{\rm pc}^2$. The 25 pixel-sized region, therefore, corresponds to an area of $\sim1.5\,{\times}\,10^3\,{\rm pc}^2$. We explore the behaviour of the [C{\sc ii}] emission for three different viewing angles and find that the trends remain unchanged. We note that the differences observed in each viewing angle are not arising from optical depth effects but rather due to the different projections of mass covered by the aforementioned area. We find that the optically thin emission holds for each different line-of-sight in these areas, as well. The most prominent feature is observed for Cluster-3. As can be seen in the second and third panel of Fig.~\ref{fig:snaps}, this cluster is formed during the second encounter (at $t{\sim}\,170\,{\rm Myr}$) and creates an H{\sc ii} region, which eventually removes the ISM that satisfies the [C{\sc ii}]-bright observational criterion (see \S\ref{sec:sims}). Thus, the [C{\sc ii}] emission of that region decreases, reflecting the trend shown in the top panel of Fig.~\ref{fig:clusters}. This indicates that local peaks of $W_{\rm CII}$ in resolved observations may provide evidence for ongoing massive cluster formation. 

\section{Origin of the [C{\sc ii}] emission}
\label{sec:origin}

\begin{figure}
    \centering
    \includegraphics[width=0.48\textwidth]{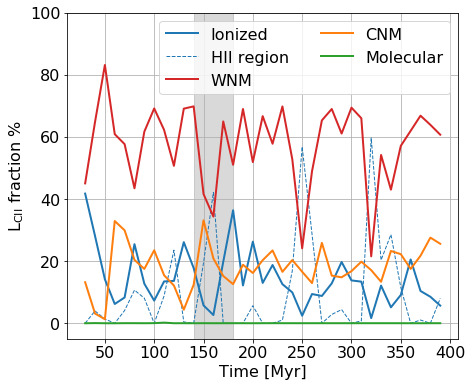}
    \caption{Percentage contribution to the total $L_{\rm CII}$ luminosity from each different ISM component. The vertical shadowed region marks the duration of the second main encounter. [C{\sc ii}] originates mainly from the WNM component at a percentage of ${\sim}58\%$. CNM contributes an approximately constant ${\sim}18\%$ at all times. A similar contribution (${\sim}14\%$) arises from the ionized material. H{\sc ii} regions contribute in general a low percentage to the total emission (${\sim}10\%$), although there are certain times where they are in an early dense evolutionary stage, thus dense, in which their contribution dominates all phases. Finally, the [C{\sc ii}] emission originating from molecular gas is always negligible.}
    \label{fig:percent}
\end{figure}

The interesting question about the origin of the [C{\sc ii}] emission has been explored by various groups both numerically and observationally. Here, we analyse the simulation outputs and we study the contribution of [C{\sc ii}] emission arising from the different ISM phases to the total emission. Each ISM phase (ionized, atomic, molecular) is identified according to the relative abundances ($\chi$) of H$^+$, H and H$_2$. In particular, the photoionized ISM (H{\sc ii} regions in which the energy of photons exceed the 13.6eV ionization potential of hydrogen) has a fixed $\chi_{\rm H^+}=0.9998$ and a gas temperature in the range $10^4<T_{\rm gas}<1.3\times10^4\,{\rm K}$. The ionized ISM (resulting from both photoionization and collisional ionization) is defined as the gas with $T_{\rm gas}>10^4\,{\rm K}$ minus the aforementioned H{\sc ii} contribution. The atomic ISM is defined as $\chi_{\rm HI}{>}2\chi_{\rm H_2}$ and the molecular ISM is defined as $\chi_{\rm HI}{<}2\chi_{\rm H_2}$. We additionally divide the atomic medium in Warm Neutral Medium (WNM; $3<\log T_{\rm gas}<4$), and Cold Neutral Medium (CNM; $\log T_{\rm gas}<3$) \citep[e.g.][]{Wolf03}. Each of the $L_{\rm CII}$ of the aforementioned four ISM phases is then compared to the total C{\sc ii} luminosity.

\begin{figure*}
    \centering
    \includegraphics[width=\textwidth]{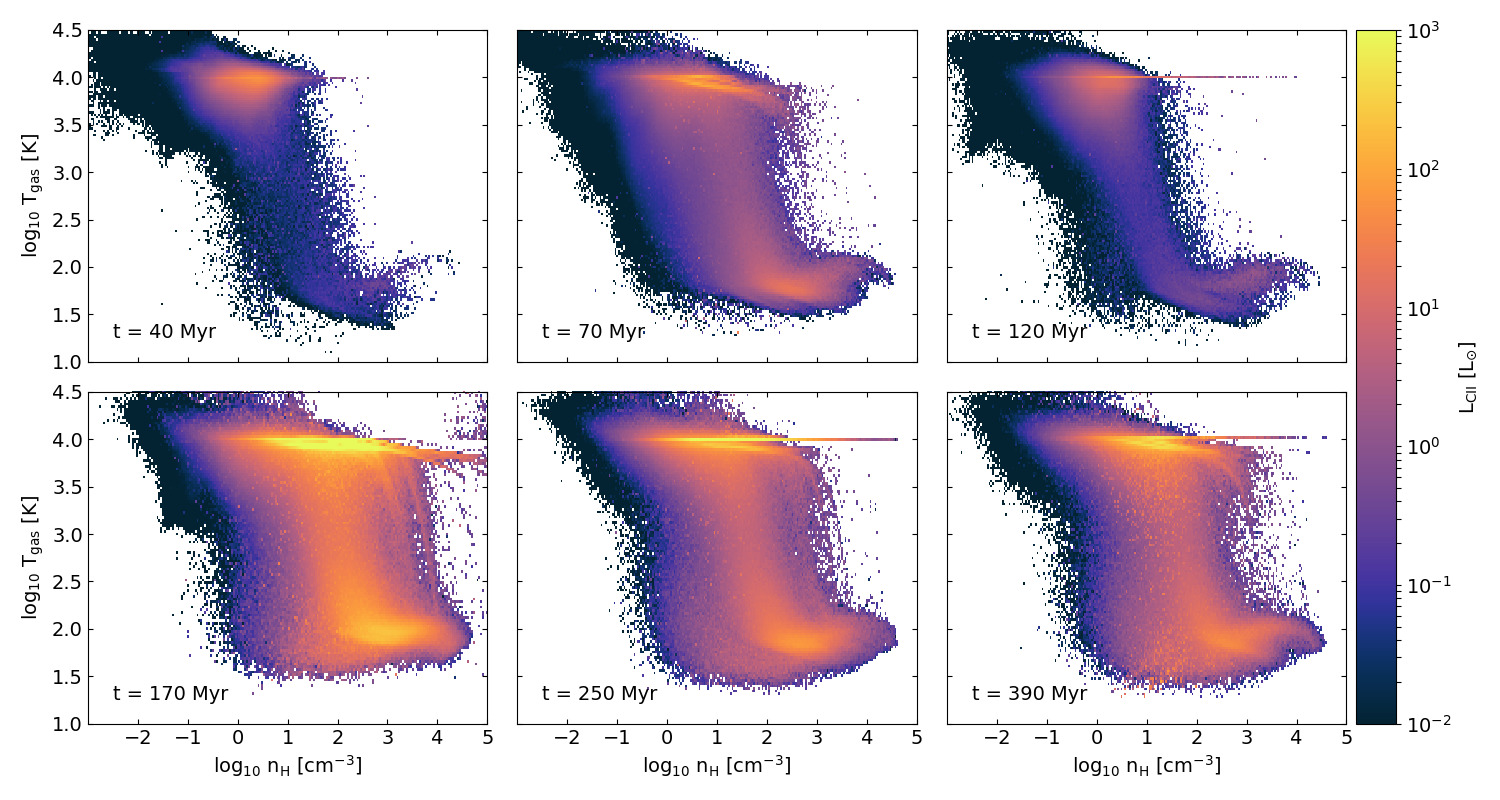}
    \caption{Phase-plots ($T_{\rm gas}$ versus total H-nucleus, $n_{\rm H}$, number density) weighted with $L_{\rm CII}$ for snapshots at $t=40,70,120,170,250$ and $390\,{\rm Myr}$. Before the second encounter (top row), $L_{\rm CII}$ originates from the WNM (see Fig.~\ref{fig:percent}). Once the second encounter occurs (bottom row), H{\sc ii} regions form and their ionized gas takes over as the main contributor to the total $L_{\rm CII}$. The horizontal straight lines at $\log T_{\rm gas}=4$ is the gas temperature of the interior of H{\sc ii} regions.}
    \label{fig:pl_Lcii}
\end{figure*}

Figure~\ref{fig:percent} shows this contribution throughout the duration of the simulation. The shaded region marks the duration of the second encounter. The emission of [C{\sc ii}] originating from WNM dominates over the corresponding emission of all other ISM phases. In particular, WNM contributes an average of ${\sim}58\%$ in agreement with previous works \citep[e.g.][]{Hu17}, whereas the contribution of the CNM is ${\sim}18\%$. The emission of the ionized gas (photoionized and collisionally ionized combined) has an average contribution of ${\sim}24\%$. Throughout the simulation, the gas that is collisionally ionized remains as the main contributor of the [C{\sc ii}] emission at this phase. Interestingly, the emission originating from H{\sc ii} regions varies substantially throughout the duration of the simulation, showing that it depends strongly on its evolutionary stage. On average, the contribution remains quite low (${\sim}5\%$). However, there are certain times e.g. at $t=160$, 250, and 320 Myr where the [C{\sc ii}] emission from H{\sc ii} regions dominate over all different ISM phases, with a contribution as high as ${\sim}50-60\%$. At early times (e.g. $t<200\,{\rm Myr}$), this sudden increase in $L_{\rm CII}$ is due to the newly formed H{\sc ii} regions which, in their early evolutionary stages, are dense and very bright in [C{\sc ii}]. Notably, such a high contribution has been recently observed in ionized regions of the inner Galaxy \citep{Lang21}. At later times after the main encounter (e.g. $t>200\,{\rm Myr}$), H{\sc ii} regions are mainly formed due to supernova explosions which create bubbles of ionized material. The contribution of $L_{\rm CII}$ originating from the molecular gas is negligible (${\sim}0.02\%$) at all times.

\begin{figure*}
    \centering
    \includegraphics[width=\textwidth]{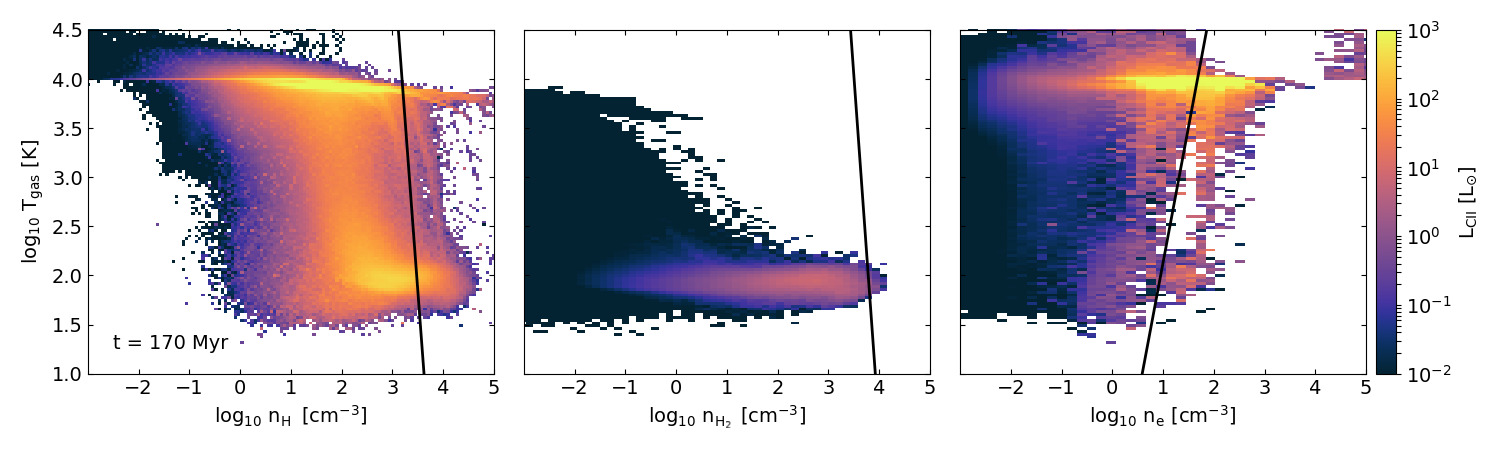}
    \caption{Phase-plots of $T_{\rm gas}$ versus the number density of the three C$^+$ colliding partners for the $t=170\,{\rm Myr}$ snapshot (see also Fig.~\ref{fig:pl_Lcii}). Here, $L_{\rm CII}$ is weighted with the abundance of each colliding partner. From left-to-right, each panel shows the number densities of atomic hydrogen (H{\sc i}), molecular hydrogen (H$_2$) and electrons (e). The black solid lines correspond to the critical density of each partner versus $T_{\rm gas}$ \citep{Gold12}. We find that collisional de-excitation is negligible.}
    \label{fig:pl_col}
\end{figure*}

\begin{figure*}
    \centering
    \includegraphics[width=0.98\linewidth]{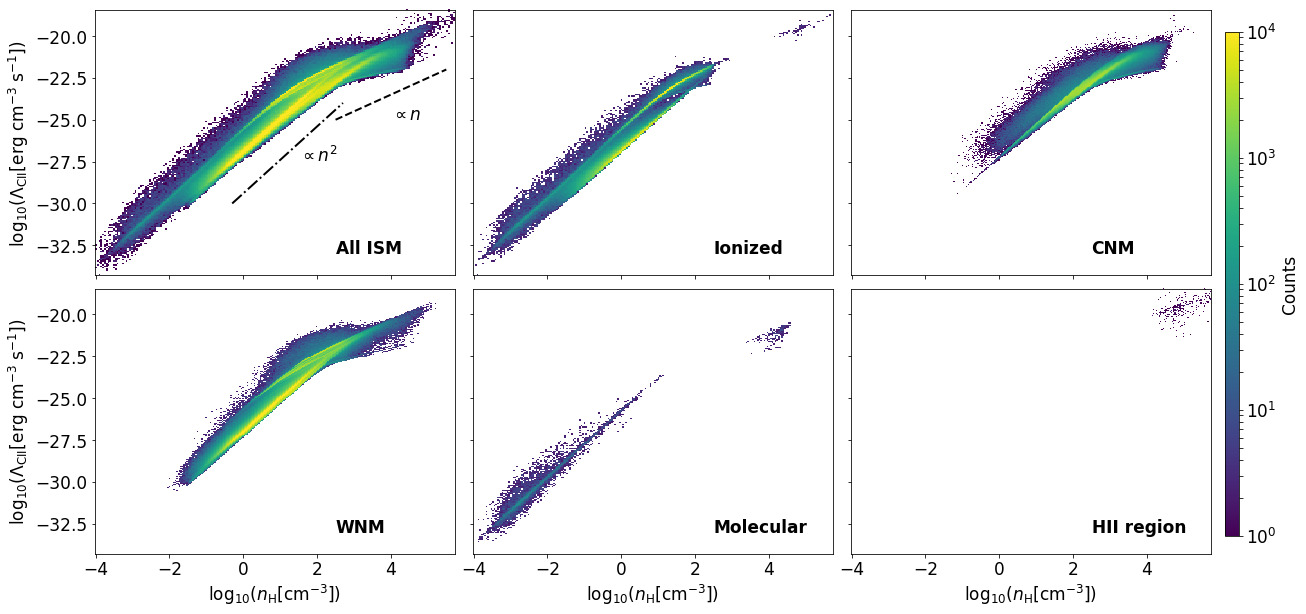}
    \caption{C{\sc ii} cooling function versus number density for the ISM gas within $1\,{\rm kpc}$ from Cluster-3 for the $t=170\,{\rm Myr}$ snapshot (see Fig.~\ref{fig:snaps}). The top left panel shows all contributing ISM gas. The dot-dashed line shows the $\propto n_{\rm H}^2$ relation while the dashed line the $\propto n_{\rm H}$ relation to guide the eye. The rest of panels show how each different ISM component (as defined in \S\ref{sec:origin}) contributes to the total $\Lambda_{\rm CII}$. In general, collisional de-excitation does not play an important role in the overall $\Lambda_{\rm CII}$ function. The color coding represents the distribution of SPH mass within the studied region.}
    \label{fig:LambdaCII}
\end{figure*}

Figure~\ref{fig:pl_Lcii} shows phase-plots (2D histograms) of gas temperature versus total number density, weighted with [C{\sc ii}] luminosity. We consider six snapshots at the times of $t=40,70,120,160,250$ and $390\,{\rm Myr}$. Comparing these panels with Fig.~\ref{fig:percent}, it can be seen that the upper bright part of the phase-plots ($\log n_{\rm H} \sim 0-2$, $\log T_{\rm gas}{>}3$), which is the WNM component of the ISM, plays the most dominant role in the origin of [C{\sc ii}]. It is interesting to note that at $t=70\,{\rm Myr}$ the bright curved rim of the WNM (starting at $\log n_{\rm H} \sim 0$, $\log T_{\rm gas} \sim 4.0$ with a declining trend as $\log n_{\rm H}$ increases) is a result of strong cooling in relatively dense and warm regions with $\log T_{\rm gas}{<}4$, which is the temperature of the ionized gas in an H{\sc ii}-region. Such strong cooling is associated with PDRs located ahead of the ionization front of the newly formed H{\sc ii} regions. This bright rim is also seen at all times during and after the second encounter, thus making PDRs a considerable contributor to the origin of [C{\sc ii}] emission. 

Before the encounter at $t=40\,{\rm Myr}$ (upper left panel of Fig.~\ref{fig:pl_Lcii}), the density of WNM component is mainly in the range of $-1<\log n_{\rm H}<1$. As the simulation progresses, the density of this ISM component increases and at the particular $t=170\,{\rm Myr}$ time (bottom left panel of Fig.~\ref{fig:pl_Lcii}), the above range extends up to $\log n_{\rm H}\sim4$. Such densities are much higher than those found locally in the Milky Way \citep[e.g.][]{Wolf95}. There are two main reasons that cause this; i) low metallicities shift the equilibrium curve to higher densities \citep[][]{Hu16}, and ii) the high FUV intensities due to feedback from cluster formation as well as supernova feedback, shift the equilibrium curve even further \citep[][]{Hu17}. In Appendix~\ref{app:equilibrium}, we additionally show mass-weighted phase-plots for the aforementioned snapshots.

Figure~\ref{fig:pl_col} shows phase-plots of the $t=170\,{\rm Myr}$ snapshot for the three colliding partners (H{\sc i}, H$_2$ and e). The C{\sc ii} luminosity is weighted with the corresponding relative abundance of each of the aforementioned colliding partners. The solid line in each panel shows the critical density of each partner as calculated by \citet{Gold12}. Gas that falls in the right-hand part of each critical density relation is collisionally de-excited. We find that at all times, $L_{\rm CII}$ associated with collisional de-excitation due to H{\sc i} and H$_2$, is negligible. The same finding applies for electrons as collision partner, except for a ${\sim}20\,{\rm Myr}$ period during the second encounter ($t{\sim}160-180\,{\rm Myr}$; see Fig.~\ref{fig:pl_col}) where the [C{\sc ii}] luminosity arising from gas with $n_{\rm e}{>}n_{\rm crit,e}$ is ${\sim}30-50\%$. This is because in this short period, compact and dense H{\sc ii} regions form, containing considerable amount of dense ionized gas. However, even during that period, collisional de-excitation still plays a minor role to the total $L_{\rm CII}$ meaning that photoelectric heating is the dominant source of [C{\sc ii}] emission at all times.

Figure \ref{fig:LambdaCII}, shows 2D histograms of the $\Lambda_{\rm CII}$ cooling function versus $n_{\rm H}$ at $t=170\,{\rm Myr}$ for the ISM gas at the inner 1~kpc from Cluster-3 (see Fig.~\ref{fig:snaps}). At this time, the emission of [C{\sc ii}] originates ${\sim}65\%$ from WNM, ${\sim}15\%$ from CNM, ${\sim}20\%$ from ionized gas, while H{\sc ii} regions and molecular gas have negligible contributions (${\lesssim}0.02\%$). As can be seen, the majority of $\Lambda_{\rm CII}$ is associated with gas with $n\lesssim10^3\,{\rm cm}^{-3}$, which is approximately the critical density for collisions with H{\sc i}. For densities lower than the aforementioned, $\Lambda_{\rm CII}$ scales as $\propto n_{\rm H}^2$ while for higher ones it scales as $\propto n_{\rm H}$ (line is thermalized). It is therefore evident that collisional de-excitation plays a minor role. In Appendix~\ref{app:rates} we present a theoretical approach as to how $\Lambda_{\rm CII}$ builds as a function of $n_{\rm H}$ and in Appendix~\ref{appB} we show 2D histograms of $\Lambda_{\rm CII}$ for the inner 0.1, 0.2 and 0.3~kpc regions from Cluster-3.

Breaking down the ISM components, it can be seen that main contributors (WNM, ionized and CNM) have a significant fraction of their $\Lambda_{\rm CII}$ associated with $n_{\rm H}\lesssim10^3\,{\rm cm}^{-3}$ gas. Similarly, the molecular component, although playing a small role, follows the same trend. It is worth mentioning that the H{\sc ii} region gas is entirely thermalized meaning that the C{\sc ii} emission that arises from this component, is a result of collisional de-excitations. This is in good agreement with the observational results of \citet{Sutt21}. H{\sc ii} regions are the places where C$^+$ is ionized to form C$^{2+}$. Given that the contribution of this ISM component is negligible, accounting for the transition between the two aforementioned ionization states of carbon can be excluded from our analysis.

In previous numerical studies, \citet{Accu17} found that ${\sim}75\%$ of the [C{\sc ii}] emission in Milky Way as well as ${\sim}60-80\%$ in galaxies of the Herschel Reference Survey, arises from their molecular regions. In molecular cloud simulations, \citet{Fran18} found that [C{\sc ii}] is primarily emitted from the cold ($T_{\rm gas}\,{\sim}\,40-65\,{\rm K}$) neutral medium with densities $n_{\rm H}{\sim}\,50-500\,{\rm cm}^{-3}$. Yet, these simulations did not include star formation and stellar feedback. In isolated dwarf galaxy simulations, \citet{Lupi20} identified the diffuse ($n_{\rm H}{\lesssim}100\,{\rm cm}^{-3}$) neutral gas to contribute most of the [C{\sc ii}] emission, while only a small fraction of [C{\sc ii}] originates from higher density gas associated with dense PDRs. They do not, however, actually show the temperature of the [C{\sc ii}] emitting gas, so the explanation that the CNM is the dominant component is only based on the density criterion. Interestingly, they do show that the typical densities of [C{\sc ii}]-bright gas are higher for lower metallicity environments. Here, we do not base our ISM definition on density cuts as the full density and temperature evolution of the gas is available to us. In this way, we identify the WNM to be the dominant source of [C{\sc ii}] emission. Following \citet{Wolf03}, our definition of WNM is based on the gas temperature. Therefore the WNM can have higher or lower densities, depending on the local balance of heating and cooling terms. The phase-plots of Fig.~\ref{fig:pl_Lcii} indicate that a $T_{\rm gas}$-based criterion is more appropriate in the case of vastly varying star formation rates, since there is warm but dense gas during the merging process, such that the WNM phase\footnote{Note also that Fig.~13 of \citet{Hu17} shows how sensitive the $L_{\rm CII}$ cumulative functions are versus density and versus $T_{\rm gas}$.} shifts to higher gas densities. Our results agree with \citet{Hu17} who also examined isolated dwarf galaxies and identified WNM to be the most [C{\sc ii}]-bright ISM phase.

\section{Relation between the [CII] line emission and the FIR emission}
\label{sec:ciidef}

\begin{figure*}
    \centering
    \includegraphics[width=0.49\textwidth]{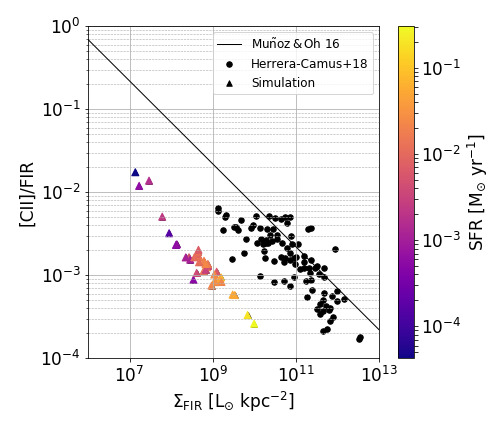}
    \includegraphics[width=0.49\textwidth]{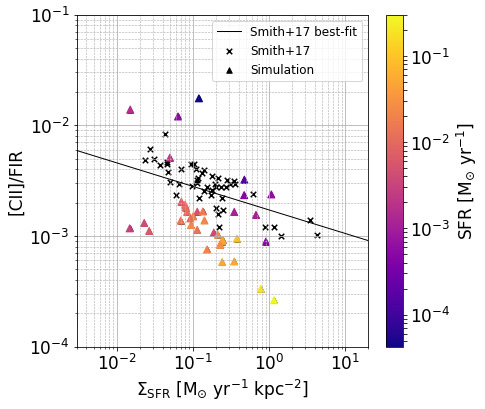}
    \caption{\textit{Left panel:} Relation of the [C{\sc ii}]/FIR ratio versus $\Sigma_{\rm FIR}$. Low [C{\sc ii}]/FIR ratios are tightly connected with high SFR values. Black circles are observations presented in \citet{Herr18}. The solid line corresponds to the \citet{Muno16} relation (Eqn.~\ref{eqn:mo}). \textit{Right panel:} Relation of the [C{\sc ii}]/FIR ratio versus $\Sigma_{\rm SFR}$. Black crosses are spatially resolved observations of the {\sc Kingfish} program presented in \citet{Smit17} and the solid line corresponds to their best-fit function (Eqn.~\ref{eqn:smith}). In both panels, triangles represent our simulation data which have been colour coded according to their SFR value.}
    \label{fig:CIIFIR}
\end{figure*}

\begin{figure*}
    \centering
    \includegraphics[width=\textwidth]{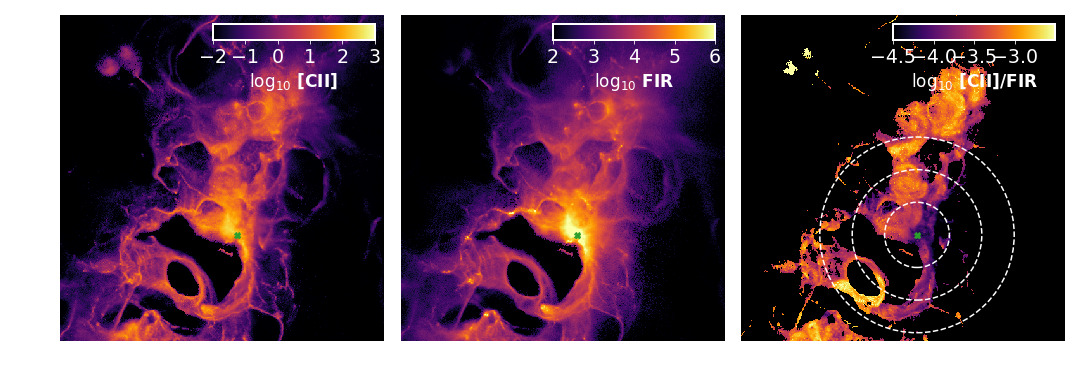}
    \caption{Zoom-in of the central region at $t=170\,{\rm Myr}$. The small green cross shows the position of Cluster-3. The left panel shows the [C{\sc ii}] luminosity, the middle panel the FIR luminosity (both in units of $\rm L_{\odot}$) and the right panel their ratio. The $L_{\rm CII}>0.5\,{\rm L}_{\odot}$ criterion has been applied to obtain this ratio. The dashed circles centered to the position of Cluster-3 show the radial distances of 0.1, 0.2 and 0.3 kpc. The inner $0.2\,{\rm kpc}$ region is [C{\sc ii}]-deficit due to the presence of strong FUV radiation field emitted from the massive Cluster-3 which increases abruptly the FIR emission while [C{\sc ii}] becomes thermally saturated.}
    \label{fig:zoom170}
\end{figure*}

The left panel of Fig.~\ref{fig:CIIFIR} shows how the [C{\sc ii}]/FIR ratio relates with $\Sigma_{\rm FIR}$, in which the observational criterion of $L_{\rm CII}>0.5\,{\rm L}_{\odot}$ has been applied. As can be seen, the ratio [C{\sc ii}]/FIR decreases with increasing $\Sigma_{\rm FIR}$. For high $L_{\rm FIR}$, this makes the ISM gas to emit more brightly in FIR in relation to [C{\sc ii}], which results in the known `[C{\sc ii}]-deficit' \citep{Malh97,Malh01,Luhm98,Luhm03,Comb18}. We plot (black squares) observations of 52 nearby galaxies ($z{<}\,0.2$) from the {\sc shining}\footnote{Survey with \textit{Herschel} of the Interstellar medium in Nearby Infrared Galaxies} sample \citep{Herr18}. 
These are not dwarf galaxies and have higher metallicities. However, we find that the [C{\sc ii}]/FIR ratio neither depends strongly on the metallicity nor does it depend strongly on the density distribution. Thus, this ratio can be compared to objects that may not necessarily satisfy the properties of a dwarf galaxy. On the other hand, $\Sigma_{\rm FIR}$ strongly depends on metallicity (assuming a linear relation between dust-to-gas and metallicity). This in turn results in a shift of the SHINING galaxies to higher $\Sigma_{\rm FIR}$ as can be seen in the left panel of Fig.~\ref{fig:CIIFIR}. Had the modelled galaxies been at solar metallicity, it would have increased the derived $\Sigma_{\rm FIR}$ by approximately one order of magnitude thereby matching with the lower end of the SHINING sample.
Nevertheless, given that a decreasing [C{\sc ii}]/FIR ratio with a comparable slope is observed in our simulations, it is interesting to explore and understand its origin.

While many different mechanisms leading to a [C{\sc ii}]-deficit medium have been proposed, we emphasize here on the effect of thermal saturation of [C{\sc ii}] emission. The effect of thermal saturation of [C{\sc ii}] leading to a [C{\sc ii}]-deficit medium was suggested by \citet{Kauf99} and studied in detail theoretically by \citet{Muno16}, with \citet{Ryba19} to provide follow-up observational evidence for its existence in dusty star-forming galaxies (with masses of $\sim10^{10}\,{\rm M}_{\odot}$) at a redshift of $z\sim3$. As \citet{Muno16} explain, the thermal saturation of [C{\sc ii}] is a direct quantum mechanical consequence of the saturation of the upper fine-structure energy state when the gas temperature exceeds the C{\sc ii} excitation temperature of $91\,{\rm K}$. Once the latter occurs, the population of the upper state cannot increase further leading to an approximately constant emissivity while the FIR dust emissivity is free to increase more. Their theoretical models lead to the expression:
\begin{eqnarray}
\label{eqn:mo}
{\rm [CII]/FIR}\simeq2.2\times10^{-3}\frac{f_{\rm CII}}{0.13}\left(\frac{\Sigma_{\rm FIR}}{10^{11}\,{\rm L}_{\odot}{\rm kpc}^{-2}}\right)^{-1/2},
\end{eqnarray}
where $f_{\rm CII}$ is the fraction of total gas traced by [C{\sc ii}]. As described in \citet{Muno16}, the value of $f_{\rm CII}=0.13$ (which is also adopted in this work) is a good estimate based on observations of Milky Way clouds and various extragalactic sources. The above relation is shown in solid black line in the left panel of Fig.~\ref{fig:CIIFIR} and its power-law is in agreement with the \citet{Herr18} observations as well as our simulations.

In general and throughout the duration of the simulation, for ${\rm SFR}>10^{-2}\,{\rm M}_{\odot}\,{\rm yr}^{-1}$ it is found that the gas is so warm that [C{\sc ii}] becomes thermally saturated. This increase in temperature is a direct consequence of the increase in FUV photoelectric  heating as a result of the high star formation activity, which eventually leads to the birth of massive star clusters. High FUV intensities are to be expected in galaxy mergers. For instance, the PDR study of \citet{Bisb14} find an average of $\langle G_0\rangle\gtrsim10^{2.5}$ in the Antennae merging system. In our simulations, the consequence of high FUV intensities for high SFRs is demonstrated with the colour bar of Fig.~\ref{fig:CIIFIR} which shows that low [C{\sc ii}]/FIR ratios are tightly connected with high values of star formation rate.

The right panel of Fig.~\ref{fig:CIIFIR} shows how the [C{\sc ii}]/FIR ratio relates with $\Sigma_{\rm SFR}$. For the latter quantity, we use the same surface, $\Sigma$, obtained from the observed area of $L_{\rm CII}>0.5\,{\rm L}_{\odot}$. The simulation points here are also colour-coded with SFR. As expected from the above discussion, the [C{\sc ii}]/FIR ratio decreases with increasing $\Sigma_{\rm SFR}$. We compare our results with observations of the {\sc Kingfish}\footnote{Key Insights on Nearby Galaxies: a Far-Infrared Survey with {\it Herschel}} program presented in \citet{Smit17}. These are spatially resolved observations of 54 nearby galaxies. Based on a galaxy sample with higher $\Sigma_{\rm SFR}$ than examined here, \citet{Smit17} find a best-fitting relation of the form:
\begin{eqnarray}
\label{eqn:smith}
{\rm [CII]/FIR}\simeq10^{-3}\left(\frac{\Sigma_{\rm SFR}}{12.7}\right)^{-1/4.7}.
\end{eqnarray}
As with the [C{\sc ii}]/FIR vs $\Sigma_{\rm FIR}$ relation, our simulations have a similar slope with the observations and the above best-fit relation. We note that Eqn.~\ref{eqn:smith} represents a single power-law fit to both local and high-redshift sources and that it can be applied when young stars provide the dominant energy source on scales greater than a few hundred parsecs \citep{Smit17}. Thus, deviations of our simulations from this power-law are to be expected.

Figure~\ref{fig:zoom170} shows a zoom-in of the central region at $t=170\,{\rm Myr}$, where three massive clusters have been formed (see also Fig.~\ref{fig:snaps}). In particular, the [C{\sc ii}] emission, FIR emission and their ratio are shown. Here, we only highlight Cluster-3 which is responsible for the strong [C{\sc ii}] and FIR emission in its immediate surroundings. In the right panel of Fig.~\ref{fig:zoom170} showing the [C{\sc ii}]/FIR ratio, the dashed circles centered on Cluster-3 show radial distances with steps of 0.1~kpc. As can be seen, the innermost part has a very low [C{\sc ii}]/FIR ratio, of the order of $10^{-4}-10^{-3}$ and it is thus `[C{\sc ii}]-deficit' when compared to the outer regions which have a [C{\sc ii}]/FIR ratio of $10^{-3}-10^{-2}$. Such a [C{\sc ii}]/FIR mapping has been observed in the central region of Orion Molecular Cloud 1 by \citet{Goic15} as well as in the wider Orion Nebula complex, recently, by \citet{Pabs21}.

\begin{figure}
    \centering
    \includegraphics[width=0.45\textwidth]{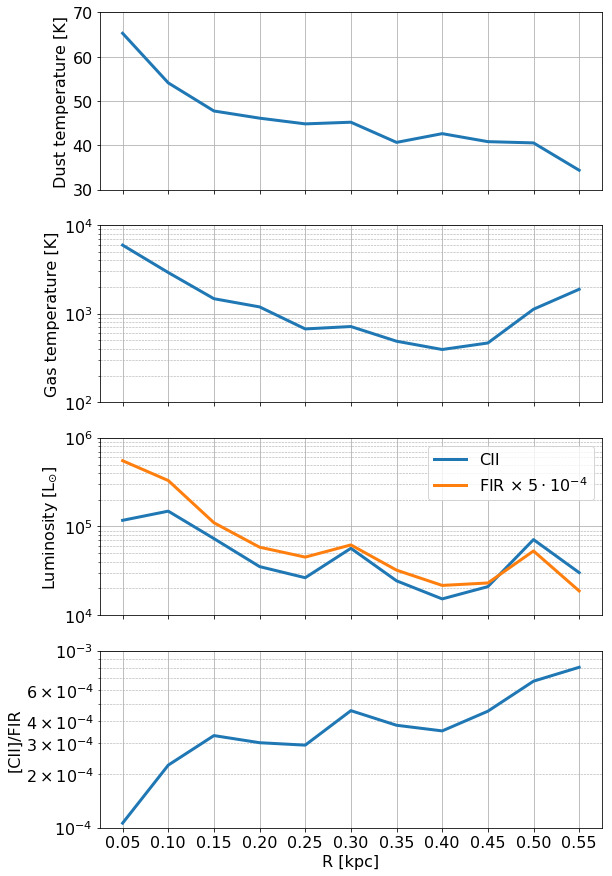}
    \caption{Density-weighted dust temperature (top panel), density-weighted gas temperature (second panel), luminosities of [C{\sc ii}] and FIR (third panel), and the [CII]/FIR ratio (bottom panel) versus the radial distance, $R$, from Cluster-3. Each quantity is averaged over shells of thickness $0.05\,{\rm kpc}$. In the third panel, the FIR luminosity is displaced downwards by a factor of $5\times10^{-4}$ to ease the comparison with [C{\sc ii}]. For small $R$, both dust and gas temperatures are high leading to high FIR and [C{\sc ii}] luminosities respectively, although the emission of [C{\sc ii}] suffers from thermal saturation while FIR is always $\propto T_{\rm d}^6$. This results in a decrease of the [C{\sc ii}]/FIR ratio ([C{\sc ii}]-deficit) as can be seen in the bottom panel. At larger $R$, the dust temperature decreases while the gas temperature remains high. This increases the [C{\sc ii}]/FIR ratio.}
    \label{fig:deficitradial}
\end{figure}

\begin{figure*}
    \centering
    \includegraphics[width=0.44\textwidth]{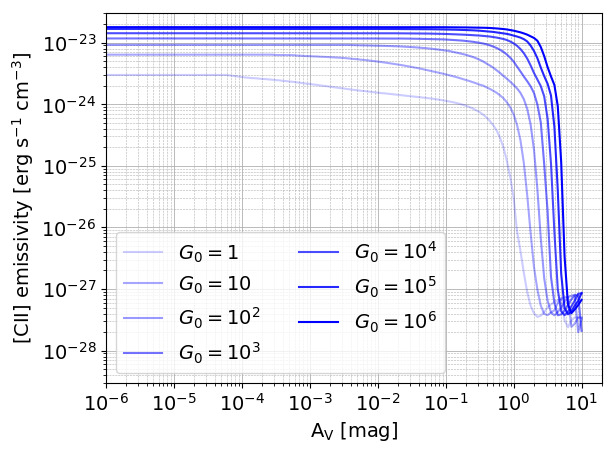}
    \includegraphics[width=0.42\textwidth]{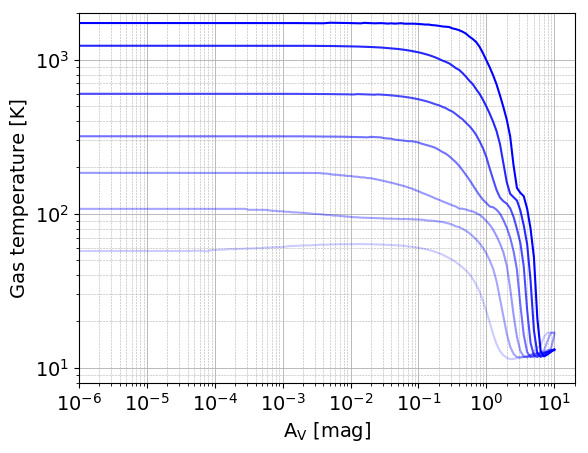}
    \includegraphics[width=0.44\textwidth]{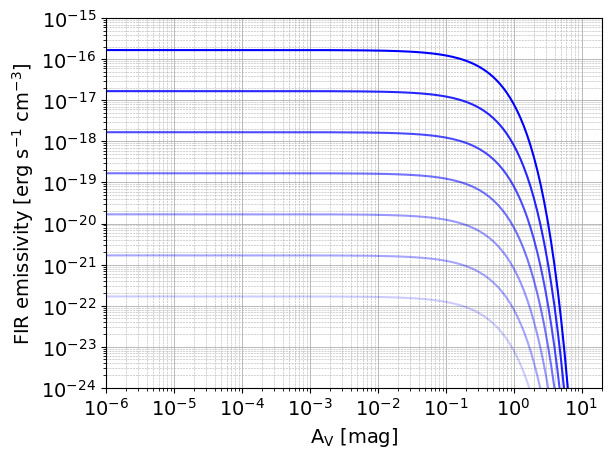}
    \includegraphics[width=0.43\textwidth]{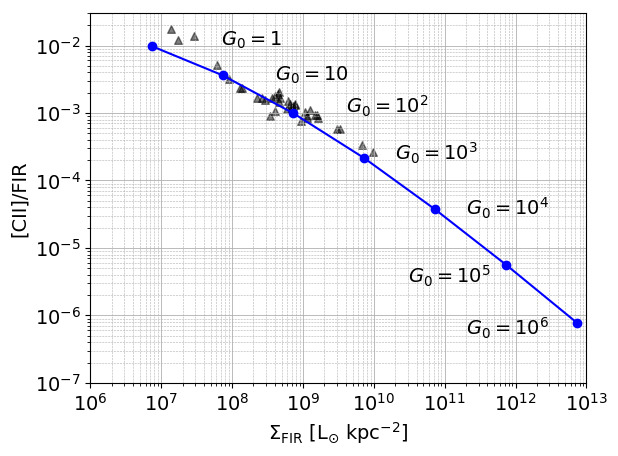}
    \caption{PDR simulation of a $n_{\rm H}=300\,{\rm cm}^{-3}$ total number density interacting with various FUV intensities in the range $G_0=1-10^6$. \textit{Top left:} emissivity of [C{\sc ii}] versus the visual extinction, $A_V$. As $G_0$ increases, [C{\sc ii}] emissivity increases from $\sim6.5\times10^{-24}$ to $\sim1.7\times10^{-23}\,{\rm erg}\,{\rm s}^{-1}\,{\rm cm}^{-3}$ at which point it saturates, approaching asymptotically a maximum value. \textit{Bottom left:} emissivity of FIR calculated assuming that the total dust cooling is equal to radiative dust heating (Eqn.~\ref{eqn:ldust2}), versus $A_V$. Under the optically thin assumption, the FIR emission is given by integrating Eqn.~(\ref{eqn:ldust2}) along the line-of-sight. \textit{Top right:} gas temperature versus $A_V$ when thermal balance has been reached. The temperature at the surface of the PDR increases from $\sim120\,{\rm K}$ to $\sim1.6\times10^3\,{\rm K}$ as $G_0$ increases. \textit{Bottom right:} [C{\sc ii}]/FIR versus $\Sigma_{\rm FIR}$ assuming optically thin emission. Due to the thermal saturation of the [C{\sc ii}] emissivity, the ratio [C{\sc ii}]/FIR decreases for $G_0>10$ in these simulations, leading to a [C{\sc ii}]-deficit medium. The gray triangles represent the simulation data as discussed in Fig.~\ref{fig:CIIFIR}.}
    \label{fig:pdr}
\end{figure*}

The above correlation can be also seen in Fig.~\ref{fig:deficitradial} in which the density-weighted dust and gas temperature as well as the luminosities of [C{\sc ii}] and FIR are plotted versus the radial distance from Cluster-3. The aforementioned quantities are averaged over shells of $0.05\,{\rm kpc}$ thickness. For $R<0.1\,{\rm kpc}$, the dust temperature is high with $T_{\rm d}>50\,{\rm K}$ which is a consequence of the very strong FUV radiation field emitted from the massive star cluster. This results in a high FIR luminosity (see Eqn.~\ref{eqn:ld}), with values up to $\sim1.7\times10^9\,{\rm L}_{\odot}$ in the $R<0.1\,{\rm kpc}$. Similarly, the high gas temperatures of $T_{\rm gas}>3\times10^3\,{\rm K}$ in that region results in a high luminosity of [C{\sc ii}] $\sim 2.7\times10^5\,{\rm L}_{\odot}$. This leads to a [C{\sc ii}]/FIR ratio of $\sim1.5\times10^{-4}$ and thus a [C{\sc ii}]-deficit gas. In the outer regions e.g. in the shell of $0.4<R<0.5\,{\rm kpc}$, the dust temperature is $\sim40\,{\rm K}$ which reduces the FIR luminosity an order of magnitude i.e. $\sim1.5\times10^8\,{\rm L}_{\odot}$. On the other hand, the gas temperature, although it is also reduced, remains much higher than the $91\,{\rm K}$ excitation temperature of [C{\sc ii}] i.e. $\sim500\,{\rm K}$. This makes the [C{\sc ii}] luminosity to decrease by a factor of $\sim3$, thus leading to a higher [C{\sc ii}]/FIR ratio. As shown in Appendix~\ref{appB}, the ISM gas immediately around the cluster is thermalized and thus grows with $\propto n_{\rm H}$. This growth cannot compensate with the $\propto T_{\rm dust}^6$ correlation of Eqn.~\ref{eqn:ld}, leading to a [C{\sc ii}]-deficit medium.

In these hydrodynamical simulations, DGR is constant in space and time. However, dust could be destroyed due to high FUV radiation fields or strong shocks \citep[e.g.][]{Drai79,Jone94,Zhuk16}. By performing the first hydrodynamical multiphase ISM simulations including dust sputtering due to SNe, \citet{Hu19} showed that DGR can decrease by ${\sim}30\%$ in the volume filling warm gas compared to that in the dense clouds. We expect that such a decrease in DGR would locally result in a lower FIR emission in regions of very high FUV intensity. At the same time, the strength of the FUV field is expected to be somehow more extended since the attenuation due to dust will be smaller and thus, $G_0$ will decrease primarily due to geometric dilution following a ${\sim}r^{-2}$ law. Considering all the above, we expect [C{\sc ii}]/FIR to locally decrease, which could result in a ``less [C{\sc ii}]-deficit" ISM gas, but the effect may be small compared to the [C{\sc ii}]/FIR value obtained from the entire simulation.

We also note that while we do not include the conversion of C{\sc ii} to C{\sc iii} in our chemical network as mentioned in \S\ref{sec:sims}, we do expect the derived [C{\sc ii}]/FIR ratio to decrease if the higher ionization states of carbon were taken into account, thereby again enhancing the [C{\sc ii}]-deficit.

\subsection{Photodissociation region calculations}

To explore the decrease in the [C{\sc ii}]/FIR ratio in greater detail, we perform PDR calculations using the publicly available {\sc 3d-pdr} photodissociation region code\footnote{https://uclchem.github.io/3dpdr} \citep{Bisb12}. The code uses the UMIST2012 database of reaction rates \citep{McEl13} and performs iterations over thermal balance by taking into account various heating and cooling processes. It calculates the abundances of species, the gas temperatures as well as the emissivities of various coolants using the Large Velocity Gradient approximation \citep{Sobo60,Cast70,deJo75}. The dust temperature due to FUV heating is calculated using the treatment of \citet{Holl91} in which $T_{\rm d}\propto G_0^{0.2}$.

In these PDR calculations, we explore the response of the [C{\sc ii}] and FIR emissivities in a one-dimensional uniform density cloud with a total H-nucleus number density of $n_{\rm H}=300\,{\rm cm}^{-3}$, as it interacts with various FUV intensities in the range $G_0=1-10^6$, normalized to the spectral shape of \citet{Drai78}. We use a subset of the UMIST2012 chemical network which contains 33 species (including e$^-$). For the purposes of this test, we also assume a cosmic-ray ionization rate of $\zeta_{\rm CR}=3\times10^{-18}\,{\rm s}^{-1}$ and metallicity of $Z=0.1\,{\rm Z}_{\odot}$ to imitate as closely as possible the adopted ISM conditions of \citetalias{Lahe20}. The cloud has a visual extinction of $A_{\rm V}=10\,{\rm mag}$ which is related to the total column density, $N_{\rm tot}$, as $A_{\rm V}=A_{\rm V,0}N_{\rm tot}(Z/Z_{\odot})$, where $A_{\rm V,0}=6.3\times10^{-22}\,{\rm mag}\,{\rm cm}^2$ \citep{Wein01,Roll07}. The size of the cloud is therefore taken to be $L\simeq170\,{\rm pc}$.

Figure~\ref{fig:pdr} illustrates the results from the PDR simulations described above. The top-left panel shows how the [C{\sc ii}] emissivity, which represents the [C{\sc ii}] cooling rate, increases for increasing $G_0$ per cloud depth. For $G_0=1$, the emissivity at the surface of the cloud is ${\sim}\,6.5\times10^{-24}\,{\rm erg}\,{\rm s}^{-1}\,{\rm cm}^{-3}$. As $G_0$ increases, the emissivity increases but becomes thermally saturated for $G_0>10^5$ at which point\footnote{This value can be analytically calculated using the expression $\Lambda_{\rm CII} = A_{ij} h \nu_{ij} n_i \beta_{ij} (S_{ij}-B_{ij})/S_{ij}$, where $A_{ij}$ is the Einstein A-coefficient, $h$ Planck's constant, $\nu_{ij}$ the [C{\sc ii}] frequency, $\beta_{ij}=1$ the escape probability at the edge of the cloud, $S_{ij}$ the source function and $B_{ij}$ the black-body function for the $2.7\,{\rm K}$ background emission. For the simulation parameters with $G_0=10^6$, {\sc 3d-pdr} outputs $n_i{\sim}6\times10^{-4}\,{\rm cm}^{-3}$ and $n_j{\sim}2\times10^{-3}\,{\rm cm}^{-3}$ ($j<i$). By replacing these values and calculating $S_{ij}$ and $B_{ij}$ accordingly, we obtain $\Lambda_{\rm CII}{\sim}1.7\times10^{-23}\,{\rm erg}\,{\rm s}^{-1}\,{\rm cm^{-3}}$.} it is ${\sim}\,1.7\times10^{-23}\,{\rm erg}\,{\rm s}^{-1}\,{\rm cm}^{-3}$ as seen in the top left panel of Fig.~\ref{fig:pdr}. Higher FUV intensities would increase the emissivity asymptotically to a maximum value, close enough to the aforementioned saturated value. On the other hand, for the assumed density of $n_{\rm H}=300\,{\rm cm}^{-3}$, the local dust cooling, corresponding to the FIR emissivity, is approximately equal to the dust heating rate due to radiation. The latter is given by the expression \citep{Glov12}
\begin{eqnarray}
\label{eqn:ldust2}
\Gamma = 5.6\times10^{-24} n_{\rm H} D' G_0\,[{\rm erg}\,{\rm cm}^{-3}\,{\rm s}^{-1}]
\end{eqnarray}
where $G_0$ is the local (attenuated) FUV intensity. Therefore, the FIR emission is given by integrating the above expression along the line-of-sight, thus $\Lambda_{\rm dust}=\int \Gamma dr$. As can be seen, the FIR emission scales linearly with the FUV intensity. The bottom-left panel shows how $\Lambda_{\rm dust}$ relates with $G_0$ per cloud depth. High FUV intensities heat up the gas, as can be seen in the top-right panel. The thermal balance calculations performed, show that for $G_0=1$ the gas temperature at the surface of the cloud is $T_{\rm gas}\,{\sim}\,120\,{\rm K}$ while for $G_0=10^6$ it is ${\sim}\,1.6\times10^3\,{\rm K}$. 

Assuming optically thin emission for both [C{\sc ii}] and FIR in this example, we integrate along the line of sight to obtain the corresponding integrated emission. This is shown in the bottom-right panel of Fig.~\ref{fig:pdr}, which correlates [C{\sc ii}]/FIR with $\Sigma_{\rm FIR}$. As $\Sigma_{\rm FIR}$ increases, [C{\sc ii}]/FIR decreases leading to a [C{\sc ii}]-deficit medium. 
Assuming a linear relation between dust-to-gas and metallicity, higher metallicities would drift the plotted curve in this panel rightwards. Here, the simulation data are also shown with gray triangle. The PDR simulation and the simulation data are in excellent agreement. 

\section{Discussion}
\label{sec:discussion}

\subsection{The star-formation rate and far-infrared luminosity relation}
\label{ssec:sfirssfr}

\citet{Kenn98} has calibrated SFR with FIR luminosity in dusty circumnuclear starbursts, providing the following relation:
\begin{eqnarray}
\label{eqn:kenn}
\frac{\rm SFR}{\rm [M_{\odot}\,yr^{-1}]}= {\rm C}\times L_{\rm FIR},
\end{eqnarray}
where ${\rm C}\simeq1.87\times10^{-10}\,{\rm L}_{\odot}$ accounts for the total IR luminosity covering the wavelength range of $3-110\mu$m \citep{Kenn12}. In the above relation, $L_{\rm FIR}$ corresponds to the total bolometric luminosity with the assumption that all FIR will emerge from dust grains heated by the interstellar FUV radiation field. In our simulations, dust heating is tightly connected with the increase of FUV radiation due to the formation of clusters and SN-feedback, so Eqn.~\ref{eqn:kenn} can be directly applied \citep[c.f.][for applying this relation to observations]{Riek09}. 

\begin{figure}
    \centering
    \includegraphics[width=0.49\textwidth]{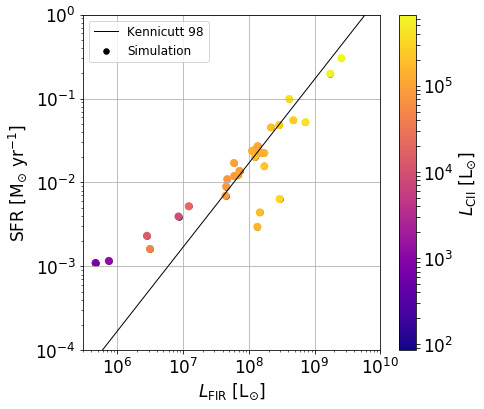}
    \caption{Relation of SFR with $L_{\rm FIR}$, colour-coded with [C{\sc ii}] luminosity. As expected, $L_{\rm CII}$ increases as SFR and $L_{\rm FIR}$ increase. The solid line corresponds to the \citet{Kenn98} relation (Eqn.~\ref{eqn:kenn}). The agreement between the simulations and the latter relation is very good.}
    \label{fig:sfirssfr}
\end{figure}

Figure~\ref{fig:sfirssfr} shows the SFR--FIR correlation for our simulations, colour coded with $L_{\rm CII}$ luminosity. As expected, the luminosity of [C{\sc ii}] increases with SFR and $L_{\rm FIR}$. The black solid line corresponds to Eqn.~\ref{eqn:kenn}. We find that our simulations are in very good agreement with the \citet{Kenn98} calibration for a broad range of $L_{\rm FIR}$ and SFR values, each one spanning approximately four orders of magnitude. Interestingly however, the agreement appears to break for lower values of SFR ($\lesssim3\times10^{-3}\,{\rm M}_{\odot}\,{\rm yr}^{-1}$). Such an effect was seen also in the recent simulations of \citet{Lahe21}. We speculate that when the UV radiation is low, there are not enough re-processed photons to produce the IR fluxes which would in turn provide reasonable estimates of the SFR predicted by Eqn.~\ref{eqn:kenn}. In this regard, \citet{Lahe21} further finds that the best agreement with the true SFR is reached with the $24\mu$m corrected UV tracers.

\subsection{The star-formation rate and [C{\sc ii}] luminosity relation}
\label{ssec:obs}

\begin{figure}
    \centering
    \includegraphics[width=0.48\textwidth]{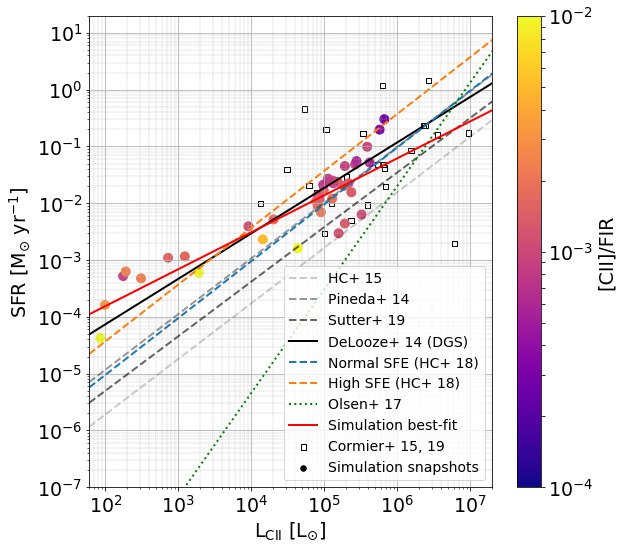}
    \caption{Comparison of simulation snapshots (circles) with observations in the ${\rm L}_{\rm CII}-{\rm SFR}$. The snapshots are colour-coded with the [C{\sc ii}]/FIR ratio. The red solid line is the best-fit from our simulation snapshots. Black and gray dashed lines represent different best-fitting relations by \citet{Herr15} (HC+ 15), \citet{Pine14}, \citet{Sutt19}. From the \citet{DeLo14} study, we plot the relation from the Dwarf Galaxy Survey (`DGS') with solid line. Furthermore, we plot the best-fit relations of galaxies with normal (blue dashed) and high (orange dashed) star formation efficiencies discussed in \citet{Herr18} (HC+ 18). In addition, we plot individual observations from DGS by \citet{Corm15,Corm19} with open black squares and with green solid line the best-fitting relation from the \citet{Olse17} simulations. We find that our simulations are in agreement with the \citet{Corm15,Corm19} observations and with the \citet{DeLo14} DGS slope. The medium becomes [C{\sc ii}]-deficit as SFR, and therefore $L_{\rm CII}$, increase.}
    \label{fig:obs}
\end{figure}

We now compare the resultant [C{\sc ii}]-SFR relation against observations found in the literature (see Appendix~\ref{app:surface} for the corresponding $\Sigma_{\rm CII}-\Sigma_{\rm SFR}$ relation). The comparison is illustrated in Fig.~\ref{fig:obs}. The simulation points are colour-coded with the [C{\sc ii}]/FIR ratio. As can be seen, the ratio decreases as both SFR and [C{\sc ii}] increase, in accordance to the discussion in \S\ref{sec:ciidef}. In the [C{\sc ii}]-SFR plane, we find that the best-fit equation representing our simulations has $m=0.65$ and $b=-5.11$ (see Eqn.~\ref{eqn:ciisfr}). 

We plot the best-fitting relations from the following four observational works. From \citet{Herr15}, who study a sample of 46 nearby star-forming galaxies from the {\it Herschel} {\sc Kingfish} survey in the absence of strong Active Galactic Nuclei (AGN). From \citet{Pine14}, who used the {\it Herschel} Galactic Observations of Terahertz C$^+$ (GOT C+) to study velocity resolved Milky Way clouds found in the Galactic plane. From \citet{Sutt19}, who studied nearby (${\lesssim}\,30\,{\rm Mpc}$) normal star-forming galaxies with no LIRGs included from {\sc Kingfish} and BtP. We also plot the \citet{DeLo14} relation of 42 dwarf galaxies from the DGS sample of \citet{Madd13}. Furthermore, we add the two $L_{\rm CII}-$SFR scalings discussed in \citet{Herr18} considering the star formation efficiency (${\rm SFE}=L_{\rm FIR}/M_{\rm mol}$, where $M_{\rm mol}$ is the molecular mass). As described in \citet{Herr18}, main-sequence, star-forming galaxies and AGNs have scalings similar to the normal SFE (blue dashed line) of the {\sc shining} survey while LINERs and (U)LIRGs have scalings similar to the high SFE (orange dashed line). During the second encounter of the collision, our simulation has a better agreement with the high SFE slope thus mimicking -even for a short period of time- the average conditions found in more massive and starburst galaxies. 

In the [C{\sc ii}]-SFR plane we find very good agreement with the slopes obtained by \citet{DeLo14} (see also Table~\ref{tab:obs}). In addition, our results compare well with the individual observations presented of DGS by \citet{Corm15,Corm19}. Furthermore, \citet{Olse17} using cosmological zoom-in simulations, presented a [C{\sc ii}]-SFR relation from 30 main-sequence galaxies at a redshift of $z{\sim}\,6$. These galaxies are of low metallicity ($Z=0.1-0.4\,{\rm Z}_{\odot}$), matching our resolved dwarf galaxy simulations, albeit the \citet{Olse17} models exhibit a higher SFR. The best-fit relation of \citet{Olse17} is shown with the green solid line (for $Z=0.1\,{\rm Z}_{\odot}$). Overall, the [C{\sc ii}] emission from the dwarf galaxy merger simulations of \citetalias{Lahe20} and their corresponding SFR values are in very good agreement with observational trends and particularly with the DGS survey \citep{Corm15,Corm19}. Notably, high-redshift galaxies with $z{\sim}5$ have been observed to satisfy the [C{\sc ii}]-SFR relation as local ($z{\sim}0$) starbursts do \citep{Herr21}. 

\section{Conclusions}
\label{sec:conclusions}

We perform [C{\sc ii}] synthetic observations in SPH simulations of low metallicity ($Z=0.1\,{\rm Z}_{\odot}$) dwarf galaxy mergers, focusing on the inner $1\,{\rm kpc}$ radius where star formation is taking place. Over time, the SFR spans more than three orders of magnitude, thus providing a useful collection of [C{\sc ii}]-SFR and FIR-SFR pairs for comparison against observations. In our analysis, we consider a lower observational limit of $L_{\rm CII}=0.5\,{\rm L}_{\odot}$, which corresponds to $\rm W(CII)\sim0.6\,{\rm K}\,{\rm km}\,{\rm s}^{-1}$, for a uniform 2D-grid resolution of $1024^2$. We find the following results:
\begin{enumerate}
    \item For systems with properties similar to the modeled ones, the emission of [C{\sc ii}] is optically thin. $L_{\rm CII}$ increases during the two merging stages, following the trend of SFR. 
    \item The simulation is in very good agreement with the \citet{Kenn98} calibration of SFR with FIR luminosity, particularly for high SFR values.
    \item We identify the Warm Neutral Medium ($3<\log T_{\rm gas}<4$, $\chi_{\rm HI}<2\chi_{\rm H2}$) to contribute an average of $\sim58\%$ to the total [C{\sc ii}] luminosity. H{\sc ii} regions contribute an average of $\sim10\%$, although when young and dense during massive star cluster formation or SNe in the form of ionized bubbles, they can become the dominant source with a contribution of $\gtrsim50\%$ for a short period of time. On the other hand, gas that is collisionally ionized may contribute an average of $\sim14\%$ to the total. Cold Neutral Medium ($\log T_{\rm gas}<3$) has a $\sim18\%$ contribution while molecular gas ($2\chi_{\rm H2}>\chi_{\rm HI}$) has negligible contribution.
    \item The ratio of [C{\sc ii}]/FIR decreases with increasing $\Sigma_{\rm FIR}$, leading to an apparent [C{\sc ii}]-deficit. We find that this occurs due to thermal saturation of [C{\sc ii}]. This is a consequence of the strong FUV heating associated with the high SFR, which increases the gas temperature to values beyond the energy separation of the ${^2}P_{3/2}-{^2}P_{1/2}$ states of [C{\sc ii}].
    The latter increases the [C{\sc ii}] emissivity to an asymptotic. On the other hand, the FIR emission increases linearly with FUV intensity.
    \item We find very good agreement with the observed trends of [C{\sc ii}]-SFR and $\Sigma_{\rm CII}-\Sigma_{\rm SFR}$ relations. Our results are in excellent agreement with the \citet{DeLo14} DGS slope and the observations of \citet{Corm15,Corm19} of the same survey. These observations best resemble the simulated systems.
\end{enumerate}

Further investigations of similar models under similar resolution will help understand the correlation of [C{\sc ii}] emission with SFR as well as with the global ISM conditions in extragalactic objects with properties similar to the simulated dwarf galaxies. In addition, different parameters in the galaxy formation and evolution model can lead to significant changes in the properties of the ISM and the star cluster formation \citep[e.g.][]{Hopk12,Buck19,Li20,Hisl22}. These can all in turn affect the star-formation rate and also the C{\sc ii} luminosity. Thus more simulations may be needed in order to have a deeper understanding of the results presented here.

\acknowledgments

The authors thank the anonymous referee for the comments and suggestions which improved the clarity of this work. TGB acknowledges support from Deutsche Forschungsgemeinschaft (DFG) grant No. 424563772 and SW thanks the DFG for funding through SFB~956 ``The conditions and impact of star formation'' (sub-project C5). SW further gratefully acknowledges funding from the European Research Council via the ERC Starting Grant RADFEEDBACK (project number 679852) under the European Community's Framework Programme FP8. We acknowledge the Leibniz Rechenzentrum (LRZ) of the Bayrische Akademie der Wissenschaften for providing super computing time on SuperMUC-NG under the grant pn72bu. TN acknowledges supported by the Excellence Cluster ORIGINS which is funded by the Deutsche Forschungsgemeinschaft (DFG, German Research Foundation) under Germany's Excellence Strategy – EXC-2094 – 390783311. UPS is supported by the Simons Foundation through a Flatiron Research Fellowship at the Center for Computational Astrophysics. The Flatiron Institute is supported by the Simons Foundation. PHJ acknowledges support by the European Research Council via ERC Consolidator Grant KETJU (no. 818930) and the support of the Academy of Finland grant 339127. The dwarf merger simulations were carried out at CSC—IT Center for Science Ltd. in Finland. 

\vspace{5mm}

\software{astropy \citep{Astropy13,Astropy18}, 
          pygad \citep{pygad},
          NumPy \citep{Harr20},
          Matplotlib \citep{Hunt07},
          RADMC-3D \citep{Dull12}, 
          SPHGal \citep{Hu14, Hu16, Hu17}, 
          RADEX \citep{vdTa07},
          3D-PDR \citep{Bisb12}
          }

\appendix

\section{RADMC-3D calculations}
\label{app:radmc3d}

We perform radiative transfer calculations in selected snapshots, using the publicly available code {\sc radmc-3d}\footnote{http://www.ita.uni-heidelberg.de/$\sim$dullemond/software/radmc-3d/} \citep{Dull12} and adopting the Large Velocity Gradient approximation \citep{Shet11}. The abundances of C$^+$, H, and H$_2$ as well as the gas temperatures and the gas velocities are taken directly from the hydrodynamical simulation. The rate coefficients for the excitation of C$^+$ and its collisions with ortho-H$_2$, para-H$_2$, H and e$^-$ are taken from the Leiden Atomic and Molecular Database\footnote{https://home.strw.leidenuniv.nl/$\sim$moldata/} \citep[LAMDA;][]{Scho05}. We considered a uniform three dimensional grid with a resolution $256^3$. The output spectra cubes have 201 channels and span $\pm200\,{\rm km}\,{\rm s}^{-1}$, giving a spectral resolution of $dv=2\,{\rm km}\,{\rm s}^{-1}$. The Doppler-catching switch is considered to account for velocity jumps between cells. We assume that the line is broadened thermally and due to microturbulence with equal contributions. To obtain the brightness temperature, we convert the {\sc radmc-3d} line intensity using the Planck function in the Rayleigh-Jeans limit. 

The computational box used in {\sc radmc-3d} has a volume of $(2\,{\rm kpc})^3$, containing the ISM of the inner 1~kpc and centered on the merging site. For each snapshot, we perform radiative transfer calculations along three different lines of sight (along $x-$, $y-$, and $z-$ axis) to account for the effects due to viewing angle. For each viewing angle, we convert the velocity integrated emission calculated with {\sc radmc-3d} to C{\sc ii} luminosity, $L_{\rm CII}$, using the expression:
\begin{eqnarray}
\label{eqn:Lcii}
L_{\rm CII} = \frac{8 \pi k_B \nu^3}{ c^3}\sum_i W_{{\rm CII},i}A_i\,[\rm L_{\odot}],
\end{eqnarray}
where $k_B$ is Boltzmann's constant, $\nu$ the rest frequency of [C{\sc ii}], $c$ the speed of light, $W_{\rm CII}$ the emission of the $i$-th pixel and $A_i$ its area. Each $2~\times~2\,{\rm kpc}^{2}$ map in the {\sc radmc-3d} calculations contains $256^2$ pixels covering equal areas. 

\section{Mass-weighted phase-plots}
\label{app:equilibrium}

\begin{figure}[h!]
    \centering
    \includegraphics[width=0.8\textwidth]{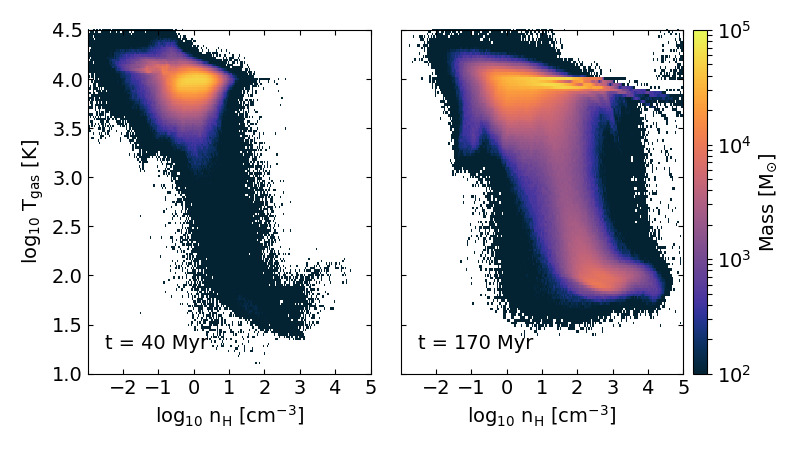}
    \caption{\bf Mass-weighted phase-plots for $t=40$ (left panel) and 170~Myr (right panel).}
    \label{fig:massweighted}
\end{figure}

Figure~\ref{fig:massweighted} shows mass-weighted plots for snapshots at $t=40$ and $t=170\,{\rm Myr}$. As can be seen, the $t=40\,{\rm Myr}$ indicates a density range of the WNM component similar to that reported for Milky Way \citep[e.g.][]{Wolf95,Wolf03}. This is also in agreement with the phase-plot presented in \citet{Lahe19} (see their Figure~1 covering a much lower density range and a much higher gas temperature range). For the $t=170\,{\rm Myr}$ snapshot, it can be seen that the origin of most of WNM mass is for densities $-1\log n_{\rm H}<3$ in the $3<\log T_{\rm gas}<4$ in temperature range.

Evidently, the emission of C{\sc ii} originates from this ISM component which, especially during the merger, contains higher densities than expected from Milky Way observations.

\section{Analytical scheme to approximate the C{\sc ii} cooling function}
\label{app:rates}

Here, we outline how to calculate the C{\sc ii} cooling rate analytically. The outlined rates are applicable for high gas temperatures. The rates of collisional de-excitation with e$^-$, H and H$_2$ as colliding partners are as follows: 
\begin{eqnarray}
R_{c,10}({\rm e}^-)&=&2.426206\times10^{-7}\left(\frac{T}{100}\right)^{0.345}\\
R_{c,10}({\rm H})&=&3.113619\times10^{-10}\left(\frac{T}{100}\right)^{0.385}\\
R_{c,10}({\rm H_2})&=&5.3\times10^{-10}T^{0.07}.
\end{eqnarray}
The rates in the above relations are measured in units of ${\rm cm}^3\,{\rm s}^{-1}$. The total de-excitation (${\cal R}_{\rm c,DEX}$) and excitation (${\cal R}_{\rm c,EX}$) rates are given respectively by the expressions:
\begin{eqnarray}
{\cal R}_{\rm c,DEX}&=&R_{c,10}({\rm H})n_{\rm H}+R_{c,10}({\rm H_2})n_{\rm H2}+R_{c,10}({\rm e})n_{\rm e}\\
{\cal R}_{\rm c,EX}&=&{\cal R}_{\rm c,DEX}\times\left(2e^{-91.25/T}\right).
\end{eqnarray}
The above rates are in units of ${\rm s}^{-1}$.

\begin{figure}[h!]
    \centering
    \includegraphics[width=0.9\textwidth]{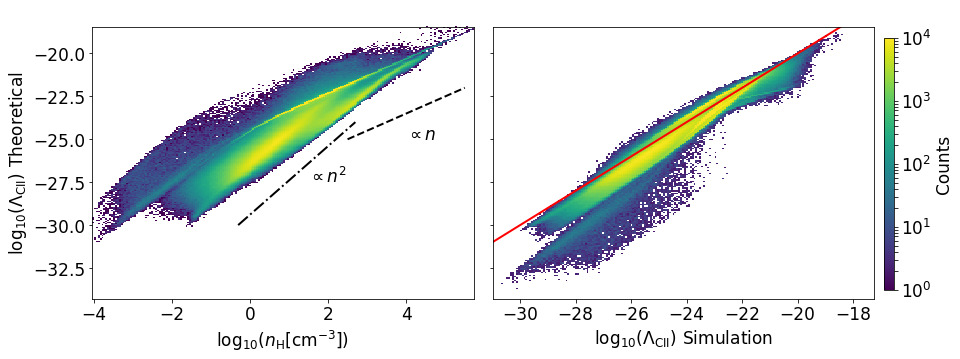}
    \caption{{\it Left panel:} Correlation of the theoretical $\Lambda_{\rm CII}$ function (Eqn.~\ref{eqn:lambdacii}) versus $n_{\rm H}$. The $\propto n_{\rm H}$ and $\propto n_{\rm H}^2$ relations are plotted for comparison. {\it Right panel:} Theoretical vs simulation CII cooling. The red solid line is the $y=x$ function to guide the eye. As can be seen the majority of $\Lambda_{\rm CII}$ function is well reproduced following the analytical expressions discussed in Appendix~\ref{app:rates}.}
    \label{fig:simtheory}
\end{figure}

The excitation temperature of C{\sc ii} is $T_{\rm CII}=h\nu/k_{\rm B}=91.25\,{\rm K}$. In case the CMB temperature is higher than  $T_{\rm CII}/5$, there will be some contribution due to stimulated emission. The contribution is negligible for the work presented in this paper. For the spontaneous emission, we make an escape probability \textit{ansatz} and use the LVG approximation. Assuming small optical depth of C{\sc ii}, the rate for spontaneous emission reduces to the corresponding Einstein-A coefficient
\begin{equation}
{\cal R}_s =A_{\rm CII} = 2.291\times 10^{-6}\; {\rm s}^{-1}.
\end{equation}

Collisional de-excitation and spontaneous emission rates are added to get the total emission rate, while the collisional excitation rate is the only contribution to the total excitation rate in case the stimulated emission is negligible. Hence, we get
\begin{eqnarray}
{\cal R}_{\rm tot, excite} = {\cal R}_{\rm c,EX}\\
{\cal R}_{\rm tot, emit} = {\cal R}_{\rm c,DEX} + {\cal R}_s. 
\end{eqnarray}

From that we define as
\begin{eqnarray}
\dot{E}_{\rm tot, excite} = \frac{{\cal R}_{\rm tot,excite}}{{\cal R}_{\rm tot,excite} + {\cal R}_{\rm tot,emit}} \times  {\cal R}_{s} \times E_{\rm CII} \\
\dot{E}_{\rm tot, emit} = \frac{{\cal R}_{\rm tot, emit}}{{\cal R}_{\rm tot,excite} + {\cal R}_{\rm tot,emit}} \times {\cal R}_{\rm CMB, EX} \times E_{\rm CII} \sim 0,
\end{eqnarray}
using $E_{\rm CII} = k_b T_{\rm CII} = h \nu_{\rm CII} = 1.25988 \times 10^{-14}\;{\rm erg}$ and assuming that $T_{\rm CMB}\ll T_{\rm CII}$.
The total cooling rate $\Lambda_{\rm CII}$ is then 
\begin{equation}
\label{eqn:lambdacii}
\Lambda_{\rm CII}= (\dot{E}_{\rm tot, excite} - \dot{E}_{\rm tot, emit}) \times {\chi'_{\rm CII}}\times n_{\rm TOT} \sim \dot{E}_{\rm tot, excite} \times {\chi'_{\rm CII}}\times n_{\rm TOT},
\end{equation}
in units of ${\rm erg}\,{\rm cm}^{-3}\,{\rm s}^{-1}$, where ${\chi'_{\rm CII}}\times n_{\rm TOT}$ is the number density of C$^+$ particles in the volume of interest.

The left panel of Fig.~\ref{fig:simtheory} plots the above equation versus the local number density, while the right one shows how it compares with the simulation result. The simulation $\Lambda_{\rm CII}$ data are taken from the snapshot at $t=170\,{\rm Myr}$ and within $1~{\rm kpc}$ from Cluster-3.

\section{$\Lambda_{\rm CII}$ cooling function around Cluster-3}
\label{appB}

\begin{figure*}[h!]
    \centering
    \includegraphics[width=\linewidth]{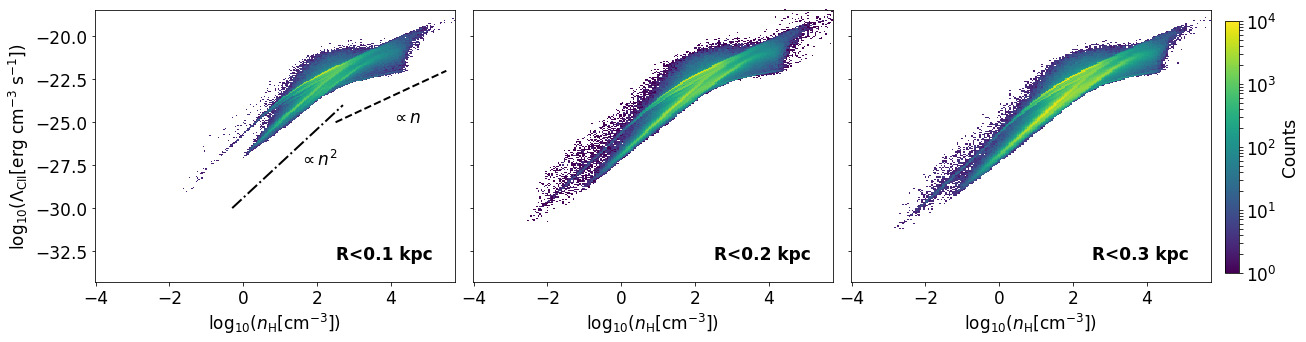}
    \caption{The C{\sc ii} cooling function versus $n_{\rm H}$ of the ISM gas within 0.1~kpc (left panel), 0.2~kpc (middle panel) and 0.3~kpc (right panel) around Cluster-3. Dot-dashed line is the $\propto n_{\rm H}^2$ and dashed line the $\propto n_{\rm H}$ relations to guide the eye. As can be seen, the ISM close Cluster-3 is by a considerable amount thermalized and thus collisionally de-excited. This causes the gas to be [C{\sc ii}]-deficit in the vicinity of Cluster-3.}
    \label{fig:appB}
\end{figure*}

Figure~\ref{fig:appB} shows 2D histograms of $\Lambda_{\rm CII}$ versus $n_{\rm H}$ within 0.1~kpc, 0.2~kpc and 0.3~kpc from Cluster-3 (see Fig.~\ref{fig:zoom170} for a visualization of the region). The ISM gas that is within 0.1~kpc has a considerable amount thermalized and thus collisionally de-excited. Since this part grows $\propto n_{\rm H}$, the emission cannot compensate with the $\propto T_{\rm dust}^6$ growth of FIR luminosity (see Eqn.~\ref{eqn:ld}), thus decreasing the [C{\sc ii}]/FIR ratio leading to a [C{\sc ii}]-deficit gas. As we increase in radial distance from Cluster-3, $\Lambda_{\rm CII}$ comes primarily form the lower density medium which grows $\propto n_{\rm H}^2$, thus increasing the [C{\sc ii}]/FIR ratio. 

\section{Effect of using a different lower observational limit for $L_{\rm CII}$}
\label{app:surface}

\begin{figure*}
    \centering
    \includegraphics[width=0.48\linewidth]{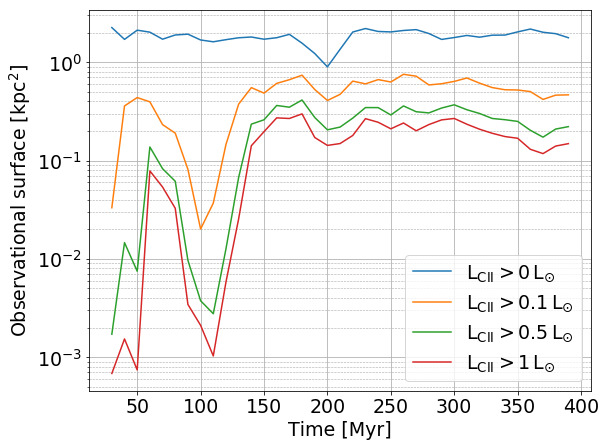}
    \includegraphics[width=0.49\linewidth]{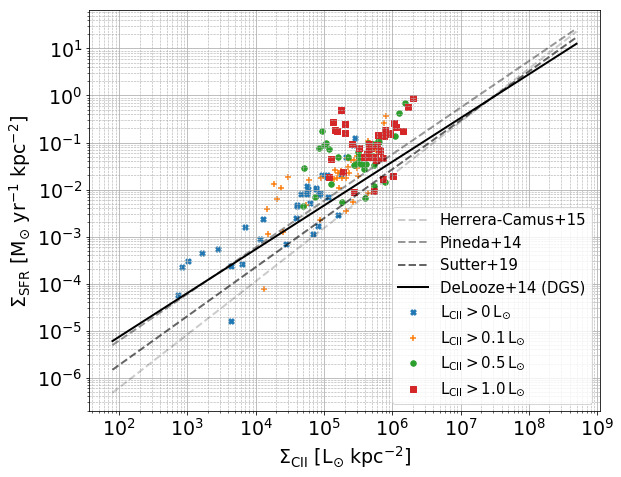}
    \includegraphics[width=0.49\linewidth]{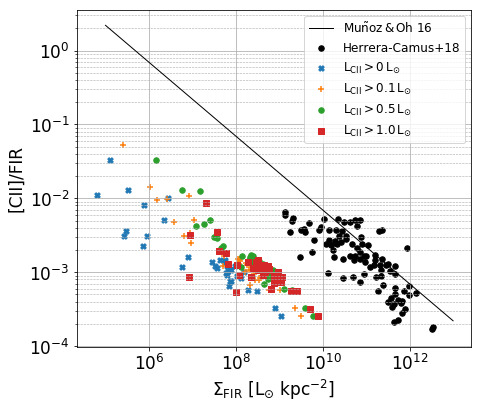}
    \includegraphics[width=0.49\linewidth]{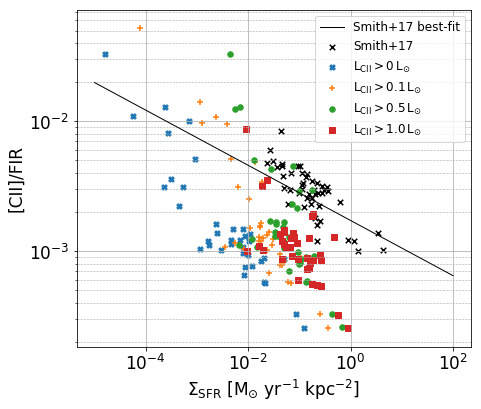}
    \caption{Effect of choosing different lower observational limit for $L_{\rm CII}$. \textit{Top left:} Time evolution of the observational surface for the four different $L_{\rm CII}$ lower limits considered. \textit{Top right:} The $\Sigma_{\rm SFR}-\Sigma_{\rm CII}$ relation. \textit{Bottom left:} The [C{\sc ii}]/FIR ratio versus $\Sigma_{\rm FIR}$. \textit{Bottom right:} The [C{\sc ii}]/FIR ratio versus $\Sigma_{\rm SFR}$. In all panels, the blue colour is for $L_{\rm CII}>0\,{\rm L}_{\odot}$, orange for $L_{\rm CII}>0.1\,{\rm L_{\odot}}$, green for $L_{\rm CII}>0.5\,{\rm L}_{\odot}$ (the one we consider in the main text) and red for $L_{\rm CII}>1\,{\rm L}_{\odot}$. As can be seen, the observational limit does not affect the trends and the overall results presented.}
    \label{fig:surfaces}
\end{figure*}

Throughout the paper, we have assumed $0.5\,{\rm L}_{\odot}$ as a lower observational limit for the C{\sc ii} luminosity. Based on this assumption, the observational surface ($\Sigma$) has been estimated which was used to calculate the $\Sigma_{\rm CII}$, $\Sigma_{\rm FIR}$ and $\Sigma_{\rm SFR}$ quantities. Here, we explore the response of the aforementioned variables if a different lower limit was adopted. In particular, we explore the cases of $L_{\rm CII}>0\,{\rm L}_{\odot}$ (all material capable of emitting [C{\sc ii}]), $L_{\rm CII}>0.1 \,{\rm L}_{\odot}$ and $>1\,{\rm L}_{\odot}$. The corresponding results are shown in Fig.~\ref{fig:surfaces}. The top left panel shows the time evolution of the observational surface when using the different $L_{\rm CII}$ limitations. The top right panel shows the response in the $\Sigma_{\rm CII}-\Sigma_{\rm SFR}$ plane. Similarly, the bottom panels show the response in the [C{\sc ii}]/FIR -- $\Sigma_{\rm FIR}$ and $\Sigma_{\rm SFR}$ planes. As can be seen, in all cases the trends and the [C{\sc ii}]/FIR ratio remain unaffected. As the lower $L_{\rm CII}$ limit increases, the observational surface decreases leading to a higher $\Sigma_{\rm SFR}$, $\Sigma_{\rm FIR}$ and $\Sigma_{\rm CII}$ values. This makes our results in the corresponding panels to drift rightwards.

\end{document}